\documentclass[a4paper,fleqn,10pt]{article}
\pdfoutput=1
\usepackage{amsmath}
\usepackage{amssymb}
\usepackage{array}
\usepackage{calc}
\usepackage{longtable}
\usepackage{multirow}
\usepackage{pstricks}
\usepackage{graphicx}
\usepackage{xspace}
\usepackage{units}

\numberwithin{equation}{section}
\usepackage{mciteplus}
\usepackage[pdfborder={0 0 0}]{hyperref}
\usepackage[format=hang,labelfont=bf,hypcap=true]{caption}
\usepackage{subcaption}
\usepackage{sectsty}
\usepackage{enumitem}
\allsectionsfont{\sffamily}
\subsubsectionfont{\mdseries\itshape\large}
\makeatletter
\DeclareRobustCommand*{\bfseries}{%
  \not@math@alphabet\bfseries\mathbf
  \fontseries\bfdefault\selectfont
  \boldmath
}
\makeatother
\let\spreprint\empty
\newcommand{\preprint}[1]{\def\spreprint{\protect#1}}
\let\sinstitute\empty
\newcommand{\institute}[1]{\def\sinstitute{\protect#1}}
\makeatletter
\renewcommand{\maketitle}{\begingroup
  \null\thispagestyle{empty}%
    \ifx\spreprint\empty
      \vskip 5ex
    \else
      \flushright\large\spreprint\vskip 10ex
    \fi
    \vskip 5ex
    \flushleft
      {\sffamily\bfseries\huge\@title}\vskip 6ex
      \@author\vskip 2ex
      \ifx\sinstitute\empty
      \else
        {\small\sinstitute}
      \fi
    \vskip 5ex
  \endgroup
}
\makeatother
\renewenvironment{abstract}{\begin{center}
  {\large\sffamily\bfseries Abstract: }
  \begin{minipage}[t]{0.75\textwidth}
}{\end{minipage}\end{center}\vskip 10ex}

\numberwithin{equation}{section}
\allowdisplaybreaks[2]

\newcommand{\LHAPDF}{L\protect\scalebox{0.8}{HAPDF}\xspace}

\newcommand{\Rivet}{R\protect\scalebox{0.8}{IVET}\xspace}

\newcommand{\Recola}{R\protect\scalebox{0.8}{ECOLA}\xspace}

\newcommand{\Collier}{C\protect\scalebox{0.8}{OLLIER}\xspace}

\newcommand{\Sherpa}{S\protect\scalebox{0.8}{HERPA}\xspace}

\long\def\symbolfootnote[#1]#2{\begingroup%
\def\thefootnote{\fnsymbol{footnote}}\footnote[#1]{#2}\endgroup}

\newcommand{\order}{\mathcal{O}}
\newcommand{\alphaS}{\alpha_s}

\newcommand{\bea}{\begin{eqnarray}}
\newcommand{\eea}{\end{eqnarray}}
\newcommand{\bi}{\begin{itemize}}
\newcommand{\ei}{\end{itemize}}
\newcommand{\hl}{\vphantom{$\int_A^B$}}

\newcommand{\shortequal}{\!\!=\!\!}

\newcommand{\deltaEW}{\ensuremath{\delta_\text{EW}}}
\newcommand{\deltaEWqq}{\ensuremath{\delta_{q\bar{q}}^\text{EW}}}
\newcommand{\deltaEWqP}{\ensuremath{\delta_{q\gamma/\bar{q}\gamma}^\text{EW}}}
\newcommand{\pT}{\ensuremath{p_\mathrm{T}}}
\newcommand{\pTmis}{\ensuremath{\displaystyle{\not}\pT}}
\newcommand{\Gmu}{\ensuremath{G_\mu}}

\newlist{myitemize}{itemize}{3}
\setlist[myitemize]{leftmargin=14em}

\newcolumntype{C}{>{\centering\arraybackslash}p{0.14\textwidth}}

\newlength{\unitcharwidth}
\hypersetup{
  pdfauthor={Marek Schoenherr},
  pdftitle={Next-to-leading order electroweak corrections to off-shell WWW production at the LHC}
}
\preprint{CERN-TH-2018-129\\MCnet-18-11}
\author{Marek Sch{\"o}nherr}
\title{Next-to-leading order electroweak corrections to\\[3mm] off-shell \texorpdfstring{$\mathrm{WWW}$}{WWW} production at the LHC}
\institute{Theoretical Physics Department, CERN, 1211 Geneva 23, Switzerland}
\begin{document}
\vspace*{10mm}
\maketitle
\vspace*{20mm}
\begin{abstract}
  Triboson processes allow for a measurement of the triple and 
  quartic couplings of the Standard Model gauge bosons, which 
  can be used to constrain anomalous gauge couplings. 
  In this paper we calculate the next-to-leading order electroweak 
  corrections to fully off-shell $W^-W^+W^+$ production, namely the 
  production of a 
  $\ell_1^-\ell_2^+\ell_3^+\bar{\nu}_{\ell_1}\nu_{\ell_2}\nu_{\ell_3}$ 
  final state with ($\ell_i=e,\mu$), including all triple, double, single 
  and non resonant topologies and interferences of diagrams with all 
  different vector boson ($W,Z,\gamma$) intermediate states. 
  We find large cancellations of the electroweak correction to the 
  $q\bar{q}$-induced channel, which includes the exchange of virtual 
  electroweak gauge bosons, and photon-induced jet radiation processes. 
  This accidental compensation is found to be strongly phase space 
  and observable dependent.
  The resulting corrections in a suitably defined fiducial region 
  thus amount to $-2.0\%$, but rise rapidly for other observables.
\end{abstract}
\newpage
\tableofcontents
\section{Introduction}
\label{sec:intro}

The precise understanding of the details of the breaking of
the electroweak symmetry of the Standard Model is one of the 
cornerstones of the LHC physics programme. 
New physics effects, which might play a role in this process, 
may conveniently be parametrised in terms of effective 
theories where heavy degrees of freedom are integrated out. 
Deviations from the Standard Model thus arise as higher-dimensional 
operators, which can lead to deviations in the triple and quartic 
couplings of the Standard Model gauge bosons. 
Deviations from the quartic gauge boson couplings in particular 
may be well measured in triboson processes, 
and $WWW$ production offers the largest cross sections.
Consquently, both ATLAS and CMS institute searches for trilepton 
production and ATLAS already used the available data to derive 
constraints on anomalous gauge couplings \cite{Aaboud:2016ftt}.

The aim to confront collider data with theorical predictions 
necessitates precise Standard Model predictions. 
Especially in the context of new physics searches all 
sources in the Standard Model known to distort spectra 
in different regions of phase space need to be accounted for. 
In this paper we calculate the next-to-leading order electroweak 
(NLO EW) corrections to off-shell $W^-W^+W^+$ production, namely to 
trilepton $\ell_1^-\ell_2^+\ell_3^+\bar{\nu}_{\ell_1}\nu_{\ell_2}\nu_{\ell_3}$ 
with $\ell_i=e,\mu$ signatures, including all triple, double, single 
and non resonant topologies and interferences of diagrams with all 
different vector boson ($W,Z,\gamma$) intermediate states. 
The QCD corrections to this process were calculated in 
\cite{Lazopoulos:2007ix,Binoth:2008kt,Campanario:2008yg} 
while the electroweak corrections were known in the literature 
for on-shell final state $W$ and $Z$ bosons, i.e.\ $WWW$ and $WZZ$ final 
states, through 
\cite{Yong-Bai:2015xna,Yong-Bai:2016sal,Dittmaier:2017bnh,Frederix:2018nkq}.
The NLO QCD corrections have also been matched to parton showers 
in \cite{Hoeche:2014rya}.

This paper is organised as follows: Sec.\ \ref{sec:trilepton} briefly 
summarises the anatomy of trilepton production processes in the context 
of off-shell triboson production and introduces the tools used in the 
calculations in this paper. Sec.\ \ref{sec:results} then presents 
the computed cross sections and correction factors to NLO EW accuracy 
for a variety of important observables before a summary is given in 
Sec.\ \ref{sec:conclusions}.

\section{Trilepton production}
\label{sec:trilepton}

\subsection{Anatomy of trilepton production processes}
\label{sec:anatomy}

In general, trilepton production may be associated with one 
neutrino ($WZ$ production), three neutrinos ($WWW$ and $WZZ$ production), 
a $b\bar{b}$-pair and three neutrinos ($t\bar{t}W$ production), 
a $b$-quark, a light quark and one neutrino ($tZj$ production), 
a $b$-quark, a light quark and three neutrinos ($tWWj$ production), 
or an even higher number of final state particles.
As this publication concerns itself with trilepton production 
in off-shell triboson processes, this category will be examined 
in detail in the following.

\begin{table}
  \centering
  \scalebox{0.79}{
    \begin{tabular}{l|l|l|l|l|l||l}
      \hl
      $\boldsymbol{\ell_1^-\ell_2^+\ell_3^+\bar{\nu}_{\ell_1}\nu_{\ell_2}\nu_{\ell_3}}$ 
      &&&&&& on-shell channels
      \\\hline\hline
      \hl
      $e^-e^+e^+\bar{\nu}_{\ell_1}\nu_{\ell_2}\nu_{\ell_3}$ 
      &&&
      $\mu^-\mu^+\mu^+\bar{\nu}_{\ell_1}\nu_{\ell_2}\nu_{\ell_3}$ 
      && 
      \\\hline
      \hl
      & $e^-e^+e^+\bar{\nu}_e\nu_e\nu_e$ 
      &&&
      $\mu^-\mu^+\mu^+\bar{\nu}_\mu\nu_\mu\nu_\mu$ 
      &&
      $WWW+WZZ$ 
      \\\hline
      \hl
      & $e^-e^+e^+\bar{\nu}_\mu\nu_\mu\nu_e$ 
      & $e^-e^+e^+\bar{\nu}_\tau\nu_\tau\nu_e$ &&
      $\mu^-\mu^+\mu^+\bar{\nu}_e\nu_e\nu_\mu$ 
      & 
      $\mu^-\mu^+\mu^+\bar{\nu}_\tau\nu_\tau\nu_\mu$&
      $\hphantom{WWW+{}}WZZ$ 
      \\\hline\hline
      \hl
      $e^-e^+\mu^+\bar{\nu}_{\ell_1}\nu_{\ell_2}\nu_{\ell_3}$ 
      &&&
      $\mu^-\mu^+e^+\bar{\nu}_{\ell_1}\nu_{\ell_2}\nu_{\ell_3}$ 
      && 
      \\\hline
      \hl
      & $e^-e^+\mu^+\bar{\nu}_e\nu_e\nu_\mu$ 
      &&&
      $\mu^-\mu^+e^+\bar{\nu}_\mu\nu_\mu\nu_e$ 
      &&
      $WWW+WZZ$ 
      \\\hline
      \hl
      & $e^-e^+\mu^+\bar{\nu}_\mu\nu_\mu\nu_\mu$ 
      &&&
      $\mu^-\mu^+e^+\bar{\nu}_e\nu_e\nu_e$ 
      &&
      $\hphantom{WWW+{}}WZZ$ 
      \\\hline
      \hl
      & $e^-e^+\mu^+\bar{\nu}_\tau\nu_\tau\nu_\mu$ 
      &&&
      $\mu^-\mu^+e^+\bar{\nu}_\tau\nu_\tau\nu_e$ 
      &&
      $\hphantom{WWW+{}}WZZ$
      \\\hline\hline
      \hl
      $e^-\mu^+\mu^+\bar{\nu}_{\ell_1}\nu_{\ell_2}\nu_{\ell_3}$ 
      &&&
      $\mu^-e^+e^+\bar{\nu}_{\ell_1}\nu_{\ell_2}\nu_{\ell_3}$ 
      &&& 
      \\\hline
      \hl
      & $e^-\mu^+\mu^+\bar{\nu}_e\nu_\mu\nu_\mu$ 
      &&&
      $\mu^-e^+e^+\bar{\nu}_\mu\nu_e\nu_e$ 
      &&
      $WWW$
      \\\hline
    \end{tabular}
  }
  \caption{
    Break-up of the contributing leptonic channels to trilepton production. 
    Channels with identical topologies and, thus, cross sections are 
    shown in the same row.
    Thus, only the processes in the first respective row will be referred to in the
    following tables and figures, while the contribution of the respective 
    identical leptonic channels to the total trilepton production cross section 
    is understood.
    The last column list the contributing on-shell $WWW$ and $WZZ$ 
    channels in this final state.
    \label{tab:channels}
  }
\end{table}

Trilepton production in the context of off-shell triboson production 
encompasses all processes of the form 
$\ell_1^-\ell_2^+\ell_3^+\bar{\nu}_{\ell_1}\nu_{\ell_2}\nu_{\ell_3}$ 
with $\ell_i=e,\mu$. 
$\tau$ lepton final states are not considered here, even though they 
may lead to similar signatures. \footnote{
  The extremely narrow width of the $\tau$ facilitates a calculation 
  factorised in production and decay of the same complexity as the 
  calculations presented here (in the limit of a massless $\tau$). 
  As only part of its energy is going into the lepton in its decay, 
  however, $\tau$-channels have generally softer spectra.
}
As indicated, only $\ell_1^-\ell_2^+\ell_3^+$ trilepton final states 
are discussed here. 
The charge conjugated final state $\ell_1^+\ell_2^-\ell_3^-$, however, 
has the same anatomy and computational complexity as the presented 
case, but the results differ due to the different parton fluxes 
involved.

$\ell_1^-\ell_2^+\ell_3^+\bar{\nu}_{\ell_1}\nu_{\ell_2}\nu_{\ell_3}$ 
with $\ell_i=e,\mu$ now involves fourteen different lepton channels, 
of which only six comprise a distinct set of Feynman diagrams. 
They are listed in Tab.\ \ref{tab:channels}. 
In its last column, this table also states the corresponding 
triboson channels in the on-shell approximation. 

\subsection{Calculational setup}
\label{sec:setup}

In this publication the combination of \Sherpa \cite{Gleisberg:2008ta,Bothmann:2016nao}
and \Recola\footnote{
  The public version 1.2 of \Recola is used.
} \cite{Actis:2012qn,Actis:2016mpe}
is used to perform all numerical calculations. 
In this combination \Sherpa provides the tree-level matrix elements, 
infrared subtraction, process management and phase-space 
integration of all contributions to all processes considered 
in this publication through its tree-level matrix element 
generator Amegic \cite{Krauss:2001iv}. 
Its inbuilt infrared subtraction \cite{Gleisberg:2007md,Schonherr:2017qcj, 
  Kallweit:2014xda,Kallweit:2015dum,Biedermann:2017yoi,Kallweit:2017khh,
  Chiesa:2017gqx,Greiner:2017mft}
is performed in the QED generalisation of the Catani-Seymour scheme 
\cite{Catani:1996vz,Dittmaier:1999mb,Catani:2002hc,Dittmaier:2008md} 
and includes the appropriate initial state mass factorisation counter 
terms. 
\Recola, on the other hand, provides the virtual corrections to all 
processes, using the \Collier library \cite {Denner:2016kdg} for the 
evaluation of its scalar and tensor integrals.
Both programs, \Sherpa and \Recola, are interfaced through the dedicated 
interface introduced in \cite{Biedermann:2017yoi}. 

\begin{table}[t!]
  \centering
  \begin{tabular}{l||r|r||r|r|r|r|r}
      & \multicolumn{2}{c||}{$N_\text{diagrams}$} & \multicolumn{5}{c}{$N_\text{channels}$} \\\cline{2-8}
      & \multicolumn{1}{c|}{B}
      & \multicolumn{1}{c||}{R}
      & \multicolumn{2}{c|}{B}
      & \multicolumn{3}{c}{R} \\\cline{2-8}
      & & 
      & \multicolumn{1}{c|}{IS} & \multicolumn{1}{c|}{FS}
      & \multicolumn{2}{c|}{IS} & \multicolumn{1}{c}{FS} \\\cline{4-8}
    Process
      & \multicolumn{1}{c|}{$q\bar{q}$} 
      & \multicolumn{1}{c||}{$q\bar{q}/q\gamma/\bar{q}\gamma$} 
      & \multicolumn{2}{c|}{$q\bar{q}$}
      & \multicolumn{1}{c|}{$q\bar{q}$}
      & \multicolumn{1}{c|}{$q\gamma/\bar{q}\gamma$}
      & \multicolumn{1}{c}{$q\bar{q}/q\gamma/\bar{q}\gamma$}\\\hline\hline
    $e^-\,e^+\,e^+\,\bar\nu_e\,\nu_e\,\nu_e$\hl
      & 614 & 5150 & 98 & 992 & 108 & 118 & 4868 \\\hline
    $e^-\,e^+\,e^+\,\bar\nu_\mu\,\nu_\mu\,\nu_e$\hl
      & 222 & 1804 & 88 & 388 & 98 & 103 & 1706 \\\hline\hline
    $e^-\,e^+\,\mu^+\,\bar\nu_e\,\nu_e\,\nu_\mu$\hl
      & 196 & 1673 & 98 & 302 & 108 & 118 & 1581 \\\hline
    $e^-\,e^+\,\mu^+\,\bar\nu_\mu\,\nu_\mu\,\nu_\mu$\hl
      & 222 & 1804 & 88 & 388 & 98 & 103 & 1706 \\\hline
    $e^-\,e^+\,\mu^+\,\bar\nu_\tau\,\nu_\tau\,\nu_\mu$\hl
      & 111 &  902 & 88 & 194 & 98 & 103 & 853\\\hline\hline
    $e^-\,\mu^+\,\mu^+\,\bar\nu_e\,\nu_\mu\,\nu_\mu$\hl
      & 170 & 1542 & 43 & 216 & 58 & 73 & 1456 \\\hline
  \end{tabular}
  \caption{
    Number of Feynman diagrams in the Born process (B) and the real 
    emission correction (R) (numbers quoted for both the $q\bar{q}$ and 
    $q\gamma/\bar{q}\gamma$ individually) and the number of associated initial 
    state (IS) and final state (FS) phase space channels, split 
    by their partonic intial states.
    \label{tab:ndiagrams-nchannels}
  }
\end{table}

All processes in the present paper are calculated including all 
triple, double, single and non-resonant topologies, as well as 
all interferences between channels with different vector boson 
intermediate states ($W,Z,\gamma$) where applicable. 
Most noteworthy here are interferences between $WWW$ and $WZZ$ 
topologies in channels with same-flavour opposite-sign lepton 
pairs.
To illustrate the computational complexity, the number of 
contributing Feynman diagrams $N_\text{diagrams}$ and associated 
channels in the phase space parametrisation $N_\text{channels}$ 
for both the Born and real emission contributions to the 
NLO EW calculation are listed in Tab.\ \ref{tab:ndiagrams-nchannels}.
The one-loop computation comprises up to eight-point loop intergrals.

In the case of on-shell $W^-W^+W^+$ production, the results 
of \cite{Dittmaier:2017bnh} were reproduced.

\nocite{Schonherr:2008av}

\section{Results}
\label{sec:results}

In this section the numerical results for the production of a 
$\ell_1^-\ell_2^+\ell_3^+\bar{\nu}_{\ell_1}\nu_{\ell_2}\nu_{\ell_3}$ 
final state ($\ell_i=e,\mu$) at next-to-leading 
order accuracy in the electroweak coupling and including all 
off-shell and interference effects at the LHC at a centre-of-mass 
energy of 13\,TeV are presented. 
All calculations are performed in the Standard Model using the 
complex mass scheme \cite{Denner:2005fg,Denner:2014zga}.
The electroweak parameters are defined in the $\Gmu$-scheme with 
the following input parameters
\begin{center}
  \begin{tabular}{rclrcl}
    $\Gmu$ &\shortequal& $1.16637\times 10^{-5}\; \text{GeV}^2$ &&& \\
    $m_W$ &\shortequal& $80.385\; \text{GeV}$       & $\Gamma_W$ &\shortequal& $2.0897\; \text{GeV}$ \\
    $m_Z$ &\shortequal& $91.1876\; \text{GeV}$      & $\Gamma_Z$ &\shortequal& $2.4955\; \text{GeV}$ \\
    $m_h$ &\shortequal& $125.0\; \text{GeV}$        & $\Gamma_h$ &\shortequal& $0.00407$\\
    $m_t$ &\shortequal& $173.2\; \text{GeV}$        & $\Gamma_t$ &\shortequal& $1.3394$\;.
  \end{tabular}
\end{center}
In this scheme, the electromagnetic coupling is defined as 
\begin{equation}
  \label{eq:defalpha}
  \begin{split}
    \alpha
    \,=&\;
      \left|
	\frac{\sqrt{2}\;\Gmu\;\mu^2_W\,\sin^2\theta_\text{w}}{\pi}
    \right|\,,
  \end{split}
\end{equation}
where the complex mass of particle $i$ and the weak mixing angle 
are defined, respectively, as
\begin{equation}
  \mu_i^2=m_i^2+\mathrm{i}\Gamma_im_i
  \qquad\text{and}\qquad
  \sin^2\theta_\text{w}=1-\frac{\mu_W^2}{\mu_Z^2}\;.
\end{equation}
The virtual amplitudes are renormalised correspondingly. 

The calculation is performed in the five-flavour scheme, i.e.\ the 
bottom quark is assumed massless and, subsequently, also considered 
as a proton constituent. 
Correspondingly, the NNPDF3.1 NLO PDF set is used \cite{Bertone:2017bme} 
including QED effects (at $\order(\alpha)$, $\order(\alphaS\alpha)$ and 
$\order(\alpha^2)$) in the parton evolution in the LUXqed scheme 
\cite{Manohar:2016nzj,Manohar:2017eqh},
interfaced through \LHAPDF \cite{Buckley:2014ana} \footnote{
  To be precise the \texttt{NNPDF31\_nlo\_as\_0118\_luxqed} PDF set 
  interfaced through \LHAPDF-6.2.1 is used.
}.
Thus, owing to the increased precision of the photon distribution in 
this new PDF set, photon induced channels appearing at NLO EW are 
determined with small PDF uncertainties.

The presented calculation is performed using the following scale 
choice
\begin{equation}\label{eq:muRF}
  \begin{split}
    \mu_R\,=\,\mu_F\,=\,3\,m_W\;.
  \end{split}
\end{equation}
Although this scale choice is not expected to provide a good description 
of the QCD dynamics of this multiscale process, the electroweak 
corrections quoted in this paper are largely independent of it.
Similarly, the CKM matrix is assumed to be diagonal.

\begin{table}[t]
  \centering
  \begin{tabular}{llc}
    Selection & Cut & Value \\\hline
    general
    \hl  & $\pT(\ell)$ & $[20\,\text{GeV},\infty)$ \\
    \hl  & $y(\ell)$ & $[-2.5,2.5]$ \\
    \hl  & $\Delta R(\ell,\ell)$ & $[0.2,\infty)$ \\\hline
    $\pTmis>20\,\text{GeV}$
    \hl  & $\Delta\phi(\pTmis,\ell\ell\ell)$ & $[\tfrac{5}{6}\pi,\pi]$ \\\hline
    1, 2 SFOS
    \hl  & $\pTmis$ & $[50\,\text{GeV},\infty)$ \\
    \hl  & $m_{\ell\ell}^\text{SFOS}$ & $[0,70\,\text{GeV}]\wedge[100\,\text{GeV},\infty)$
  \end{tabular}
  \caption{
    Definition of the fiducial region. 
    All selection cuts are applied to dressed leptons, defined with 
    $\Delta R_\text{dress}=0.1$.
    \label{tab:cuts}
  }
\end{table}

In the following analysis a set of acceptance cuts, adapted and 
idealised from \cite{Aaboud:2016ftt} and listed in Tab.\ \ref{tab:cuts}, 
defines the inclusive fiducial phase space. 
All selections are applied to dressed leptons, defined 
through recombining the bare lepton four-momentum with 
the momenta of all photons in a cone with radius 
$\Delta R_\text{dress}=0.1$ around it. 
The fiducial region requires exactly three such dressed leptons that 
have a transverse momentum larger than 20\,GeV and lie within 
a rapidity interval of $-2.5$ to 2.5. 
Further, any pair of leptons is to be separated by at least 
$\Delta R(\ell,\ell)=0.2$. 
Additional jet activity is suppressed by requiring near 
back-to-back kinematics of the leptonic system and the 
missing transverse momentum carried by the neutrinos. 
If the missing transverse momentum is larger than 20\,GeV, 
where it can be measured reasonably accurate in the LHC 
experiments, the azimuthal separation of the three-lepton-system 
and the missing transverse momentum must not be smaller 
than $\tfrac{5}{6}\,\pi$. 
This is done to both minimise backgrounds from $t\bar{t}W$, $tZj$ 
and $tWWj$ production as well as to control large higher-order 
QCD corrections that originate in additional jet emissions 
and the opening of the respective new channels. 
At the same-time, they serve to reduce the impact of the 
photon-induced real emission corrections which were found 
to be large \cite{Dittmaier:2017bnh}.

As the three different production channels, exhibiting 
no, one or two lepton pairs of the same flavour and opposite sign 
of its charge (0, 1 or 2 SFOS), are differently affected by 
their main background, $WZ\to 3\ell+\nu$ production, the set 
of applied cuts somewhat differs between them. 
In the case that no SFOS lepton pair is present, no $WZ$ 
background is present eihter and no further cut is applied. 
In the 1 and 2 SFOS case, however, a region around each 
$Z$ boson resonance is excluded by demanding the invariant 
mass of every SFOS lepton pair to lie outside the interval
$(70\,\text{GeV},100\,\text{GeV})$.
A supplementary cut on the missing transverse momentum, 
requiring it to be larger than 50\,GeV, further suppresses 
the $WZ$ backgrounds. 
The complete analysis has been implemented in \Rivet 
\cite{Buckley:2010ar}.

In the following, we first present inclusive cross sections 
before moving on to differential distributions in Sec.\ 
\ref{sec:results:diff-xs}.
In both cases, we detail the respective electroweak corrections, 
\deltaEW, and its disassembly into the respective partonic 
channels, \deltaEWqq and \deltaEWqP, defined through
\begin{equation}
  \label{eq:ew-corr}
  \begin{split}
    \langle O\rangle^\text{NLO EW}
    \,=\;
      \langle O\rangle^\text{LO}\times\deltaEW
    \,=\;
      \langle O\rangle^\text{LO}\times
      \left[
	\;
	\deltaEWqq + \deltaEWqP\vphantom{\int}
	\;
      \right]\;.
  \end{split}
\end{equation}
In this way, we are able to attribute correctly the respective 
sizes of genuine (electro)weak corrections (typically negative) 
and the (positive) photon induced QED-type corrections, 
despite their extensive \emph{accidental} 
cancellation, already observed in the on-shell case 
\cite{Dittmaier:2017bnh}.

\subsection{Inclusive cross sections}
\label{sec:results:inc-xs}

In this section inclusive cross sections and their NLO 
EW corrections, divided according to quark- and photon-induced 
channels, are presented. 
Further, the combined 
$\ell_1^-\ell_2^+\ell_3^+\bar{\nu}_{\ell_1}\nu_{\ell_2}\nu_{\ell_3}$ 
($\ell_i=e,\mu$) 
cross section is shown as well its subdivision into individual 
leptonic channels. 
Here, only the unique channels are displayed and all identical 
channels can directly inferred by the reader using the identification 
of Tab.\ \ref{tab:channels}. 
For brevity, it is understood that in cases where the neutrino species 
does not appear in the label of the process they are summed over.

\begin{table}[t!]
  \centering
  \begin{tabular}{ll||C|C|C|C}
  \hl  & & 
  \multicolumn{4}{c}{inclusive} \\
  \cline{3-6}
  \hl  & &
  LO [fb] & \deltaEW & \deltaEWqq & \deltaEWqP \\
  \hline\hline
  $\boldsymbol{\ell^-\ell^+\ell^+}$\hl & & 
  $0.4209$  & $\hspace*{\unitcharwidth}-2.0\,\%$  & $\hspace*{\unitcharwidth}-5.2\,\%$  & $\phantom{-}\hspace*{\unitcharwidth}3.2\,\%$  \\
  \hline\hline
  $e^-e^+e^+$\hl & & 
  $0.0212$  & $\hspace*{\unitcharwidth}-3.4\,\%$  & $\hspace*{\unitcharwidth}-7.1\,\%$  & $\phantom{-}\hspace*{\unitcharwidth}3.6\,\%$  \\
  \hline
  & $e^-e^+e^+\nu_e\nu_e\bar\nu_e$\hl & 
  $0.0206$  & $\hspace*{\unitcharwidth}-3.4\,\%$  & $\hspace*{\unitcharwidth}-7.0\,\%$  & $\phantom{-}\hspace*{\unitcharwidth}3.6\,\%$  \\
  \hline
  & $e^-e^+e^+\nu_e\nu_{\mu/\tau}\bar\nu_{\mu/\tau}$\hl & 
  $0.0006$  & $\hspace*{\unitcharwidth}-5.4\,\%$  & $\hspace*{\unitcharwidth}-9.5\,\%$  & $\phantom{-}\hspace*{\unitcharwidth}4.1\,\%$  \\
  \hline\hline
  $e^-e^+\mu^+$\hl & & 
  $0.0938$  & $\hspace*{\unitcharwidth}-1.4\,\%$  & $\hspace*{\unitcharwidth}-5.4\,\%$  & $\phantom{-}\hspace*{\unitcharwidth}4.1\,\%$  \\
  \hline
  & $e^-e^+\mu^+\nu_\mu\nu_e\bar\nu_e$\hl & 
  $0.0924$  & $\hspace*{\unitcharwidth}-1.4\,\%$  & $\hspace*{\unitcharwidth}-5.4\,\%$  & $\phantom{-}\hspace*{\unitcharwidth}4.1\,\%$  \\
  \hline
  & $e^-e^+\mu^+\nu_\mu\nu_\mu\bar\nu_\mu$\hl & 
  $0.0007$  & $\hspace*{\unitcharwidth}-2.9\,\%$  & $\hspace*{\unitcharwidth}-6.1\,\%$  & $\phantom{-}\hspace*{\unitcharwidth}3.2\,\%$  \\
  \hline
  & $e^-e^+\mu^+\nu_\mu\nu_\tau\bar\nu_\tau$\hl & 
  $0.0007$  & $\hspace*{\unitcharwidth}-2.7\,\%$  & $\hspace*{\unitcharwidth}-6.2\,\%$  & $\phantom{-}\hspace*{\unitcharwidth}3.5\,\%$  \\
  \hline\hline
  $e^-\mu^+\mu^+$\hl & & 
  $0.0955$  & $\hspace*{\unitcharwidth}-2.2\,\%$  & $\hspace*{\unitcharwidth}-4.6\,\%$  & $\phantom{-}\hspace*{\unitcharwidth}2.4\,\%$  \\
  \hline
  & $e^-\mu^+\mu^+\nu_\mu\nu_\mu\bar\nu_e$\hl & 
  $0.0955$  & $\hspace*{\unitcharwidth}-2.2\,\%$  & $\hspace*{\unitcharwidth}-4.6\,\%$  & $\phantom{-}\hspace*{\unitcharwidth}2.4\,\%$  \\
  \hline
  \hline
\end{tabular}
  \caption{
    Inclusive cross section and electroweak corrections in the 
    fiducial region defined by the cuts of Table \ref{tab:cuts}. 
    The remaining processes can be obtained by replacing 
    $e^\pm\leftrightarrow\mu^\pm$ throughout, cf.\ Tab.\ \ref{tab:channels}.
    \label{tab:inc-xs}
  }
\end{table}

Tab.\ \ref{tab:inc-xs} now displays the inclusive cross section 
in the complete fiducial region defined above, cf.\ Tab.\ \ref{tab:cuts}. 
The 0 and 1 SFOS lepton pair channels each contribute roughly 
45\% of the cross section, while the 2 SFOS channel (also at a 
disadvantage due to its final state symmetry factor of $\tfrac{1}{2}$), 
only contributes about 10\%. 
Similarly interesting is a decomposition in terms of associated 
processes in the on-shell approximation, cf.\ Tab.\ \ref{tab:channels}. 
While the pure $WWW$ channel only contributes 45\%, the channels 
with contributions from both the $WWW$ and $ZZZ$ channels contribute 
53\%. 
The pure $WZZ$ channels, due to the exclusion of resonant $Z\to\ell\ell$ 
processes through the fiducial cuts, only contribute 1\%. 
It is thus reasonable to conclude that the channels which contain 
$WWW$ and $WZZ$ resonances are dominated by their $WWW$ topologies. 
Hence, the fiducial cuts project the trilepton final state well 
onto $WWW$ production structures.

The size of the electroweak correction varies between different leptonic 
channels, ranging from $-1.4\%$ for $e^-e^+e^+\bar\nu_e\nu_e\nu_e$ 
to $-5.4\%$ for $e^-e^+e^+\bar\nu_e\nu_e\nu_e$. 
Generally, the pure $WZZ$ channels receive larger corrections 
than those with a $WWW$ component. 
As observed before, the complete electroweak correction suffers from 
\emph{accidental} cancellations between the genuine (electro)weak 
corrections in the $q\bar{q}$ channel and the QED-type real jet radiation 
corrections in the $q\gamma/\bar{q}\gamma$ channels. 
While the former ranges from $-4.6\%$ to $-9.5\%$, the latter compensates 
with $+2.4\%$ to $+4.1\%$. 
The largest cancellations are observed in the $e^-e^+e^+\bar\nu_e\nu_e\nu_e$ 
channel, resulting in the smallest overall correction. 
Of course, the precise size of the photon-induced corrections are 
strongly dependent on the precise value and variable used to 
suppress jet activity.
A less restrictive criterion can lead to much larger positive contributions, 
cf.\ \cite{Dittmaier:2017bnh}.

\begin{table}[t!]
  \centering
  \begin{tabular}{ll||C|C|C|C}
  \hl  & & 
  \multicolumn{4}{c}{$m(3\ell)>500\,\text{GeV}$} \\
  \cline{3-6}
  \hl  & &
  LO [fb] & \deltaEW & \deltaEWqq & \deltaEWqP \\
  \hline\hline
  $\boldsymbol{\ell^-\ell^+\ell^+}$\hl & & 
  $0.0338$  & $\hspace*{\unitcharwidth}-7.7\,\%$  & $-16.3\,\%$  & $\phantom{-}\hspace*{\unitcharwidth}8.6\,\%$  \\
  \hline\hline
  $e^-e^+e^+$\hl & & 
  $0.0031$  & $-10.1\,\%$  & $-18.3\,\%$  & $\phantom{-}\hspace*{\unitcharwidth}8.2\,\%$  \\
  \hline
  & $e^-e^+e^+\nu_e\nu_e\bar\nu_e$\hl & 
  $0.0029$  & $\hspace*{\unitcharwidth}-9.9\,\%$  & $-18.3\,\%$  & $\phantom{-}\hspace*{\unitcharwidth}8.3\,\%$  \\
  \hline
  & $e^-e^+e^+\nu_e\nu_{\mu/\tau}\bar\nu_{\mu/\tau}$\hl & 
  $0.0001$  & $-13.4\,\%$  & $-19.8\,\%$  & $\phantom{-}\hspace*{\unitcharwidth}6.4\,\%$  \\
  \hline\hline
  $e^-e^+\mu^+$\hl & & 
  $0.0081$  & $\hspace*{\unitcharwidth}-6.8\,\%$  & $-16.6\,\%$  & $\phantom{-}9.8\,\%$  \\
  \hline
  & $e^-e^+\mu^+\nu_\mu\nu_e\bar\nu_e$\hl & 
  $0.0079$  & $\hspace*{\unitcharwidth}-6.5\,\%$  & $-16.5\,\%$  & $\phantom{-}10.0\,\%$  \\
  \hline
  & $e^-e^+\mu^+\nu_\mu\nu_\mu\bar\nu_\mu$\hl & 
  $0.0001$  & $-11.9\,\%$  & $-18.0\,\%$  & $\phantom{-}\hspace*{\unitcharwidth}6.1\,\%$  \\
  \hline
  & $e^-e^+\mu^+\nu_\mu\nu_\tau\bar\nu_\tau$\hl & 
  $0.0001$  & $-11.2\,\%$  & $-17.8\,\%$  & $\phantom{-}\hspace*{\unitcharwidth}6.6\,\%$  \\
  \hline\hline
  $e^-\mu^+\mu^+$\hl & & 
  $0.0057$  & $\hspace*{\unitcharwidth}-7.7\,\%$  & $-14.8\,\%$  & $\phantom{-}\hspace*{\unitcharwidth}7.0\,\%$  \\
  \hline
  & $e^-\mu^+\mu^+\nu_\mu\nu_\mu\bar\nu_e$\hl & 
  $0.0057$  & $\hspace*{\unitcharwidth}-7.7\,\%$  & $-14.8\,\%$  & $\phantom{-}\hspace*{\unitcharwidth}7.0\,\%$  \\
  \hline
  \hline
\end{tabular}

  \caption{
    Cross section and electroweak corrections in the high trilepton 
    invariant mass region defined by adding the requirement 
    $m(3\ell)>500\,\text{GeV}$ to the fiducial cuts of Table \ref{tab:cuts}. 
    The remaining processes can be obtained by replacing 
    $e^\pm\leftrightarrow\mu^\pm$ throughout, cf.\ Tab.\ \ref{tab:channels}.
    \label{tab:highmlll-xs}
  }
\end{table}

Tab.\ \ref{tab:highmlll-xs} now additionally requires a minimal trilepton 
invariant mass of 500\,GeV. 
The total trilepton cross section thus drops to 8\% of the inclusive value. 
The importance of the 2 SFOS lepton pair channel almost doubles 
to 18\% while the 1 SFOS lepton pair channel grows only marginally to 
48\%. 
The 0 SFOS channel consequently drops to 34\%. 
Again, taken apart by corresponding on-shell channels, the 
pure $WZZ$ channels increase slightly to 2\%, leaving 98\% 
to $WWW$ dominated channels.

The electroweak corrections, through the interplay of $q\bar{q}$ 
and $q\gamma/\bar{q}\gamma$ channels, show an interesting behaviour 
in the different channels. 
While \deltaEWqq takes values from $-14.8\%$ to $-19.8\%$ 
(generally slightly larger for processes containing $WZZ$ topologies), 
\deltaEWqP ranges from $+6.1\%$ to $+10.0\%$ (generally slightly smaller 
for processes without $WWW$ topologies), resulting 
in combined corrections that are substantially smaller. 
Again, the precise details of this accidental cancellation 
depend on the precise form of any jet veto applied.

\begin{table}[t!]
  \centering
  \begin{tabular}{ll||C|C|C|C}
  \hl  & & 
  \multicolumn{4}{c}{$\pTmis>200\,\text{GeV}$} \\
  \cline{3-6}
  \hl  & &
  LO [fb] & \deltaEW & \deltaEWqq & \deltaEWqP \\
  \hline\hline
  $\boldsymbol{\ell^-\ell^+\ell^+}$\hl & & 
  $0.0097$  & $\hspace*{\unitcharwidth}-3.4\,\%$  & $-20.7\,\%$  & $\phantom{-}17.3\,\%$  \\
  \hline\hline
  $e^-e^+e^+$\hl & & 
  $0.0009$  & $\hspace*{\unitcharwidth}-3.1\,\%$  & $-23.3\,\%$  & $\phantom{-}20.2\,\%$  \\
  \hline
  & $e^-e^+e^+\nu_e\nu_e\bar\nu_e$\hl & 
  $0.0007$  & $\hspace*{\unitcharwidth}-1.8\,\%$  & $-23.9\,\%$  & $\phantom{-}22.2\,\%$  \\
  \hline
  & $e^-e^+e^+\nu_e\nu_{\mu/\tau}\bar\nu_{\mu/\tau}$\hl & 
  $0.0001$  & $-11.0\,\%$  & $-20.1\,\%$  & $\phantom{-}\hspace*{\unitcharwidth}9.1\,\%$  \\
  \hline\hline
  $e^-e^+\mu^+$\hl & & 
  $0.0027$  & $\hspace*{\unitcharwidth}-3.5\,\%$  & $-19.9\,\%$  & $\phantom{-}16.4\,\%$  \\
  \hline
  & $e^-e^+\mu^+\nu_\mu\nu_e\bar\nu_e$\hl & 
  $0.0025$  & $\hspace*{\unitcharwidth}-3.1\,\%$  & $-20.2\,\%$  & $\phantom{-}17.1\,\%$  \\
  \hline
  & $e^-e^+\mu^+\nu_\mu\nu_\mu\bar\nu_\mu$\hl & 
  $0.0001$  & $\hspace*{\unitcharwidth}-8.1\,\%$  & $-16.5\,\%$  & $\phantom{-}\hspace*{\unitcharwidth}8.4\,\%$  \\
  \hline
  & $e^-e^+\mu^+\nu_\mu\nu_\tau\bar\nu_\tau$\hl & 
  $0.0001$  & $\hspace*{\unitcharwidth}-8.0\,\%$  & $-16.7\,\%$  & $\phantom{-}\hspace*{\unitcharwidth}8.7\,\%$  \\
  \hline\hline
  $e^-\mu^+\mu^+$\hl & & 
  $0.0013$  & $\hspace*{\unitcharwidth}-3.3\,\%$  & $-20.6\,\%$  & $\phantom{-}17.3\,\%$  \\
  \hline
  & $e^-\mu^+\mu^+\nu_\mu\nu_\mu\bar\nu_e$\hl & 
  $0.0013$  & $\hspace*{\unitcharwidth}-3.3\,\%$  & $-20.6\,\%$  & $\phantom{-}17.3\,\%$  \\
  \hline
  \hline
\end{tabular}
  \caption{
    Cross section and electroweak corrections in the high missing transverse 
    momentum region defined by adding the requirement 
    $\pTmis>200\,\text{GeV}$ to the fiducial cuts of Table \ref{tab:cuts}. 
    The remaining processes can be obtained by replacing 
    $e^\pm\leftrightarrow\mu^\pm$ throughout, cf.\ Tab.\ \ref{tab:channels}.
    \label{tab:highmet-xs}
  }
\end{table}

Finally, Tab.\ \ref{tab:highmet-xs} shows the inclusive cross sections 
after requiring a missing transverse momentum of more than 200\,GeV in 
addition to the fiducial cuts of Tab.\ \ref{tab:cuts}. 
Now, the total cross section is reduced to 2.3\% of the inclusive 
value and also its compostion into lepton flavour channels changed 
markedly.
While the 0 SFOS lepton pair channel is reduced to 26\%, the 2 SFOS 
lepton pair channel is raised to 18\%. 
The 1 SFOS channel, with 56\%, also contributes more than in the 
other selections. 
The contributions of the pure $WZZ$ channels rises to about 6\%, 
leaving the total cross section still being dominated by $WWW$ 
channels.

The electroweak corrections, at least on the level of the individual 
contributions are also larger than in the other two selections.
Again, \deltaEWqP is larger in lepton channels with $WWW$ topologies, 
ranging here from $+17.1\%$ to $22.2\%$, whereas the pure $WZZ$ channels 
exhibit only half that correction. 
On the contrary, \deltaEWqq is relatively uniform with $-16.5\%$ to 
$-23.9\%$. 
The total correction is thus, again, accidentally small with $-3.4\%$.

\subsection{Differential distributions}
\label{sec:results:diff-xs}

We now turn the focus of our discussion on differential distributions. 
Each of figures discussed in the following will consist of four 
panels. 
In clockwise direction they are:
\begin{itemize}
  \item[a)] The upper left panel displays the absolute trilepton 
            cross section at leading and next-to-leading order in 
            the electroweak coupling. 
	    The upper of its two ratio plots shows the combined 
	    electroweak correction \deltaEW as well as its two components 
	    \deltaEWqq and \deltaEWqP.
	    The lower ratio plot details the composition of the sample 
	    into the 0, 1 and 2 SFOS lepton pair channels.
  \item[b)] The upper right panel presents the 0 SFOS lepton pair channel. 
	    As this channel has only one subchannel, it is not further 
	    subdivided. 
	    The ratio displays the composition and size of its 
	    electroweak correction.
  \item[c)] The lower right panel details the 1 SFOS lepton pair channel, 
	    and its three distinct subchannels. 
	    The three ratio plots show the size and composition of the 
	    respective electroweak corrections for each subchannel.
  \item[d)] The lower left panel shows the 2 SFOS lepton pair channel, 
	    and its two distinct subchannels.
	    Both ratio plots demonstrate the size and 
	    composition of their respective electroweak corrections.
\end{itemize}

\begin{figure}[p]
  \centering
  \begin{minipage}{0.47\textwidth}
    \includegraphics[width=\textwidth]{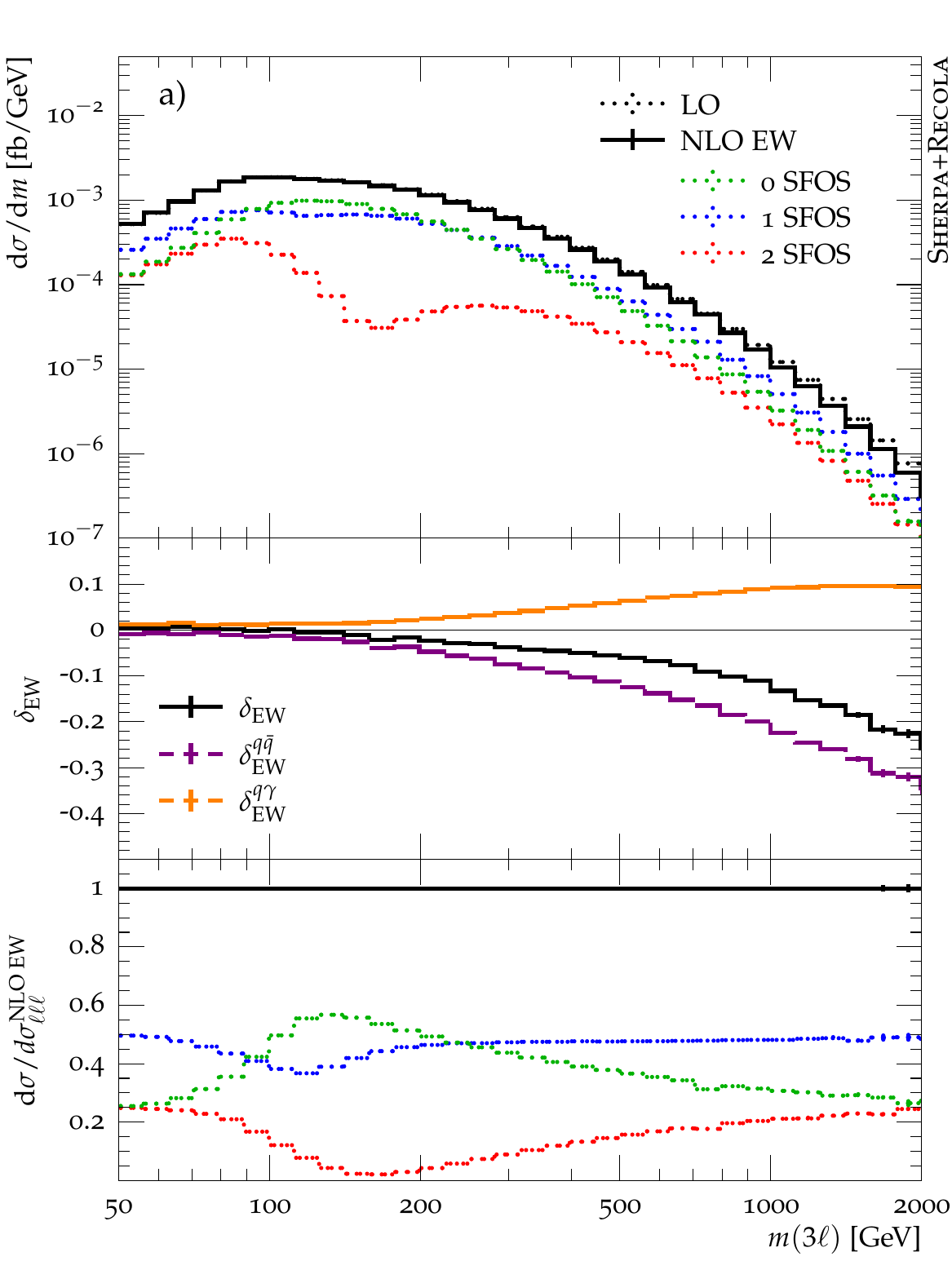}
    \includegraphics[width=\textwidth]{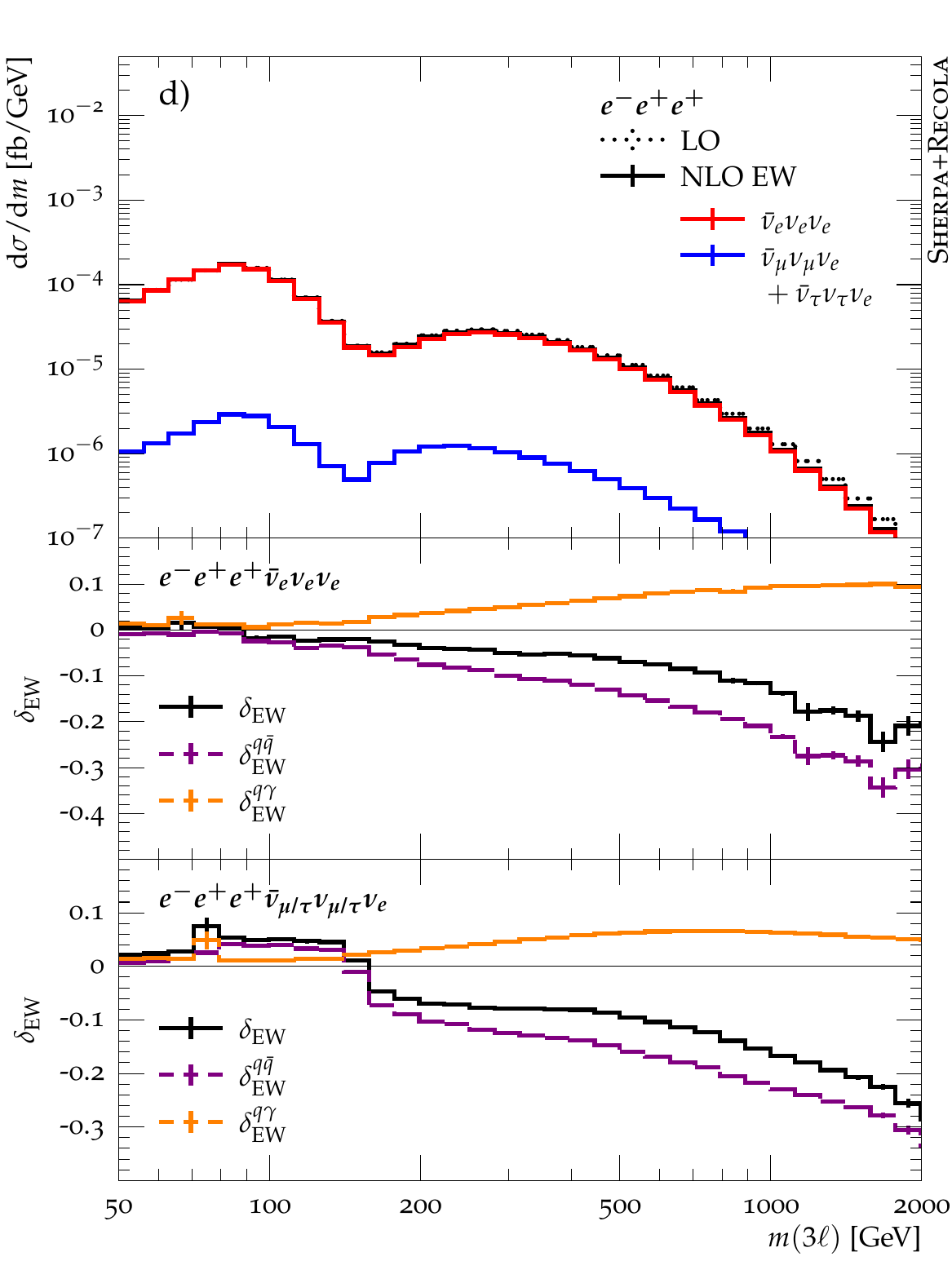}
  \end{minipage}
  \hfill
  \begin{minipage}{0.47\textwidth}
    \includegraphics[width=\textwidth]{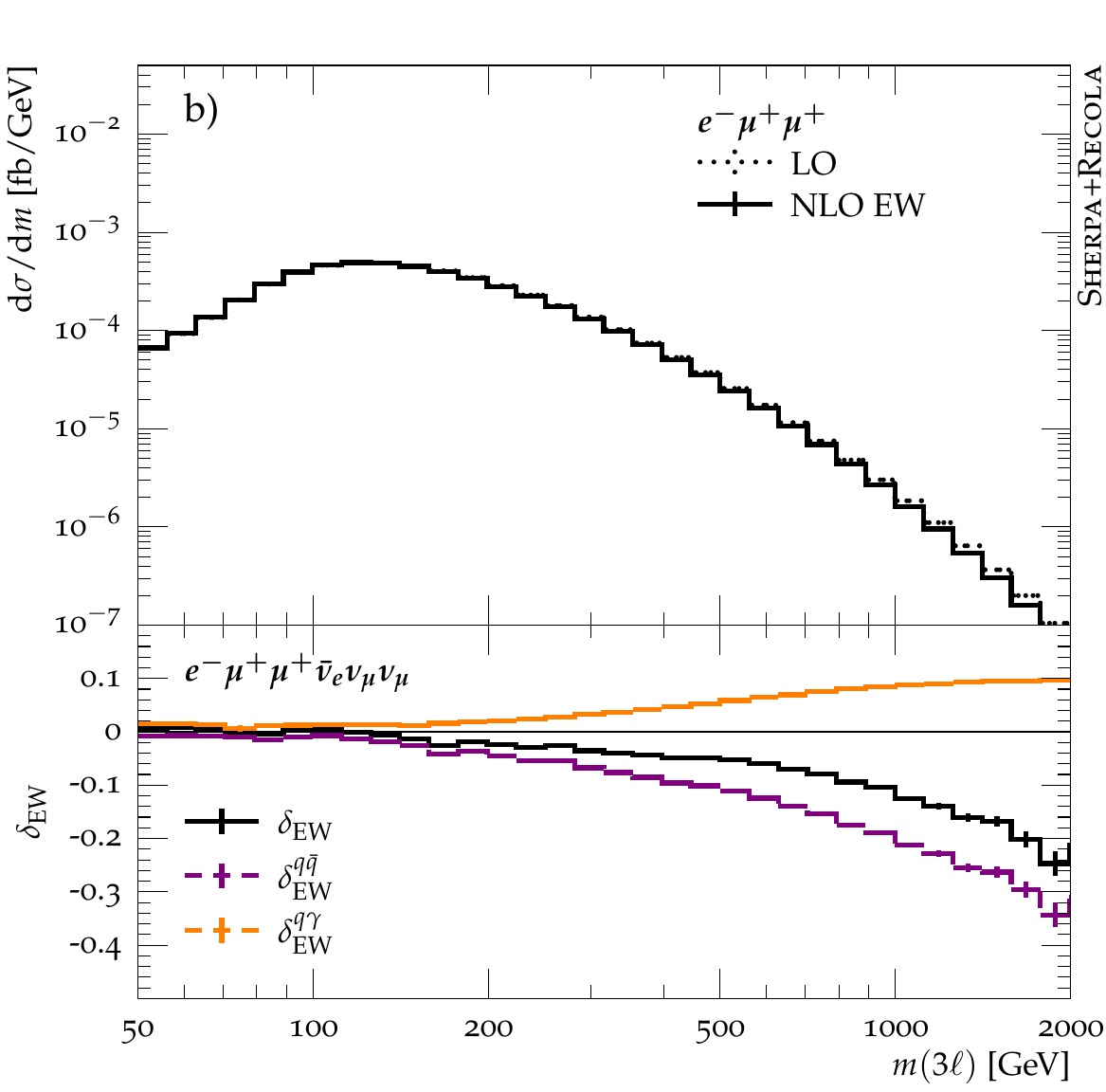}
    \includegraphics[width=\textwidth]{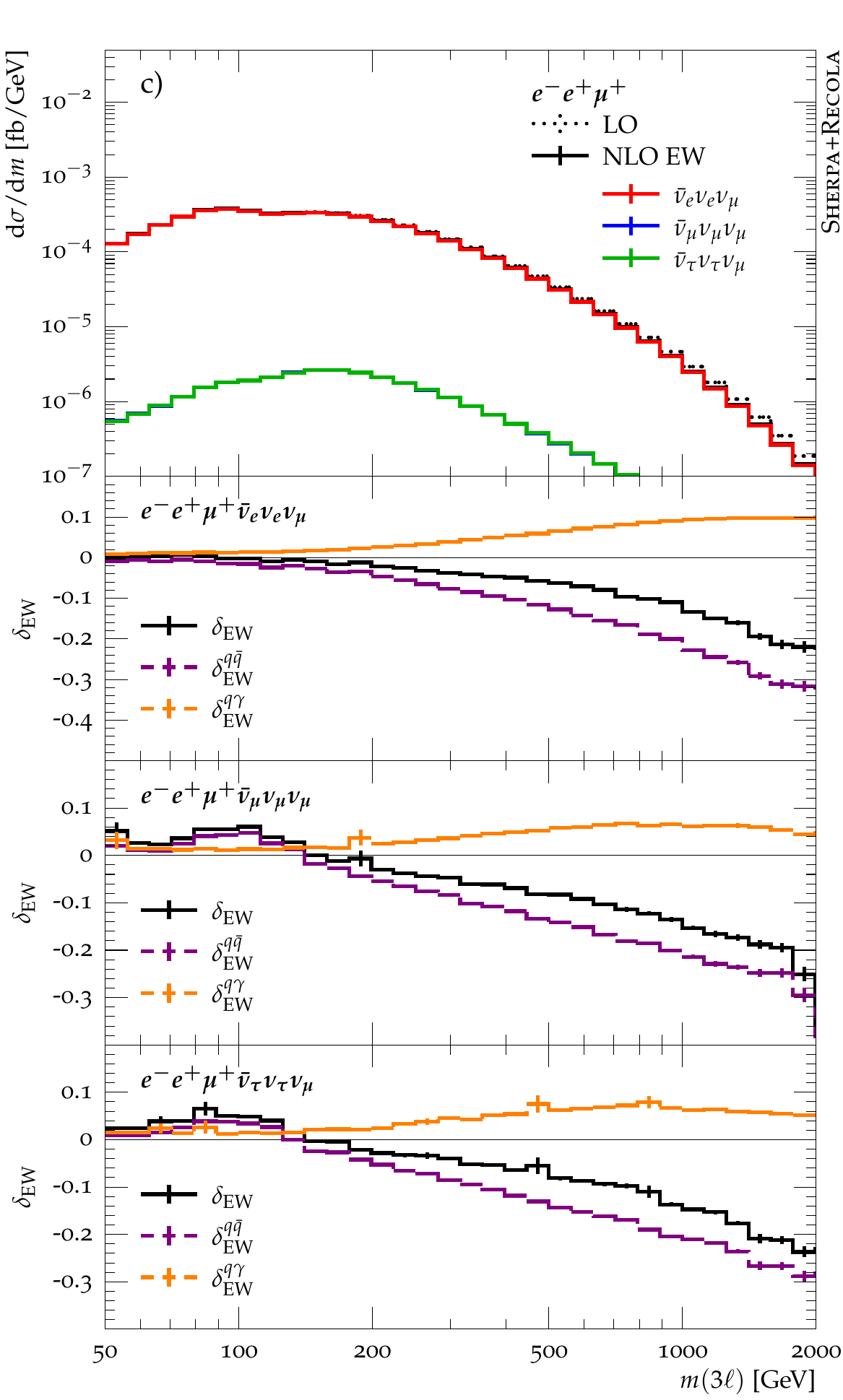}
  \end{minipage}
  \caption{
    Electroweak corrections to the trilepton invariant mass distribution. 
    \label{fig:m_3l}
  }
\end{figure}

Fig.\ \ref{fig:m_3l} shows the electroweak corrections for the trilepton 
invariant mass distribution, both for the complete trilepton final 
state and broken down for each distinct lepton channel. 
The first thing to realise is that the large contribution 
of the 0 SFOS lepton pair channel as well as the subdominance of the 
2 SFOS lepton pair channel originate in the application of the 
rejection of resonant $Z\to\ell\ell$ topologies in the 1 and 2 SFOS 
lepton pair channels. 
Especially the 2 SFOS lepton channel is suppressed over wide phase 
space regions. 
Secondly, the contribution of the photon-induced real emission correction 
saturates at high $m_{3\ell}$ (and even decreases beyond the plotted 
range). 
Consequently, the combined electroweak corrections \deltaEW\ 
exhibits its familiar EW Sudakov behaviour in that region. 
The individual lepton channels all show very similar behaviour, 
with the onset of the decreasing \deltaEWqP\ happening much 
earlier for the subdominant pure $WZZ$ processes. 
It is also worth noting that in these channels \deltaEWqq\ 
turns positive at small $m_{3\ell}$, far below the on-shell 
threshold (at $\tfrac{1}{2}\,(m_W+2m_Z)$).

\begin{figure}[p]
  \centering
  \begin{minipage}{0.47\textwidth}
    \includegraphics[width=\textwidth]{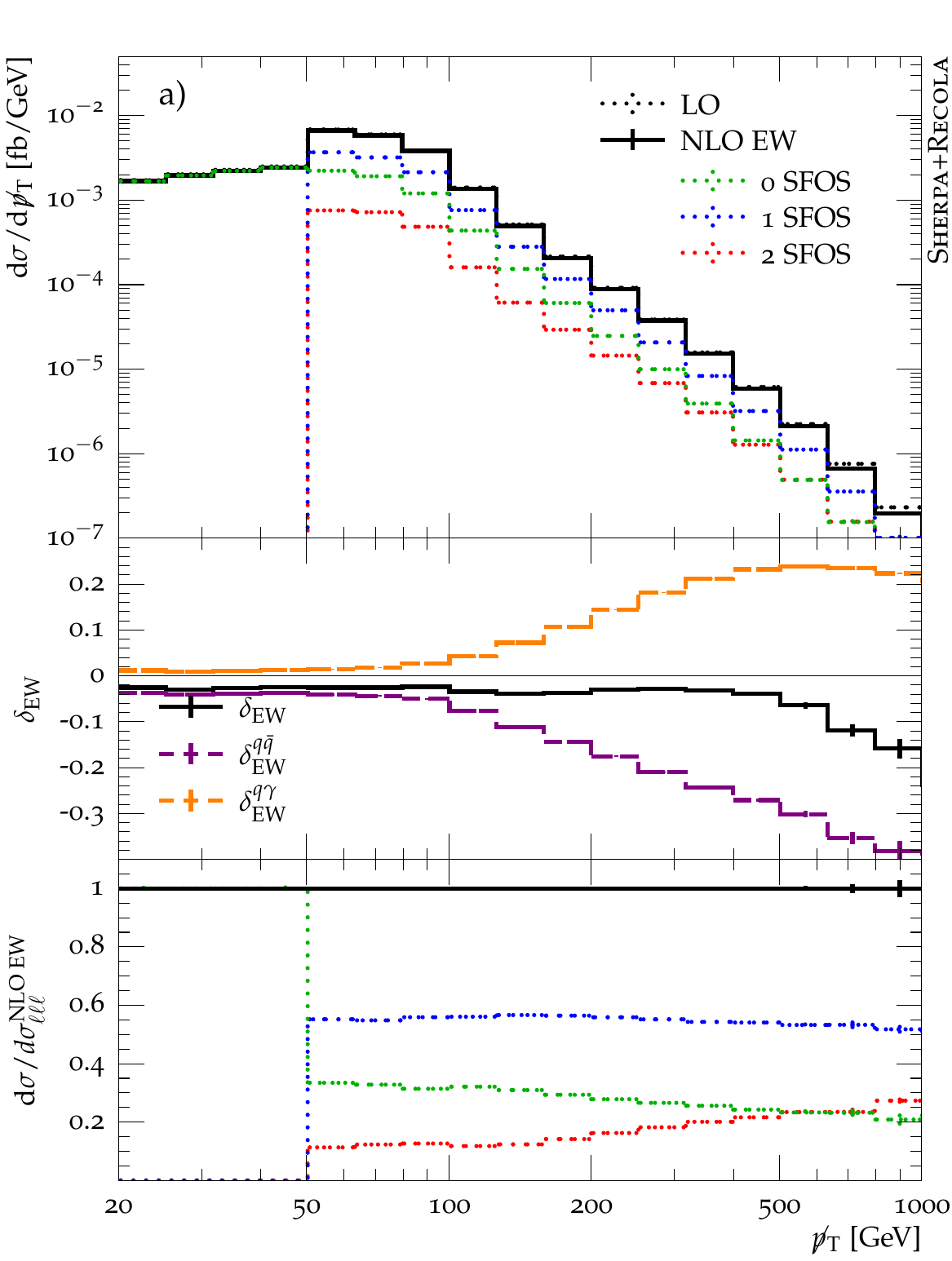}
    \includegraphics[width=\textwidth]{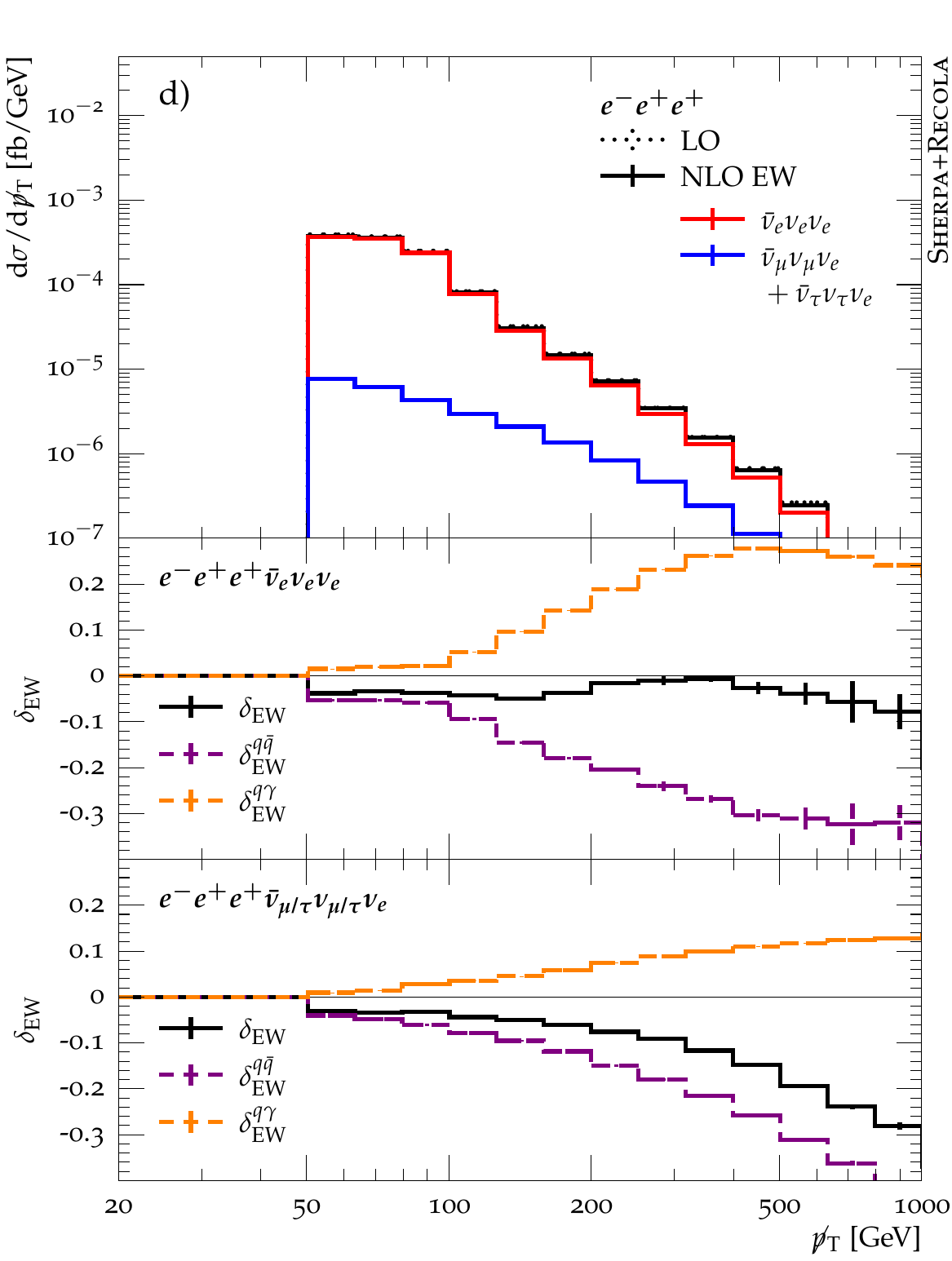}
  \end{minipage}
  \hfill
  \begin{minipage}{0.47\textwidth}
    \includegraphics[width=\textwidth]{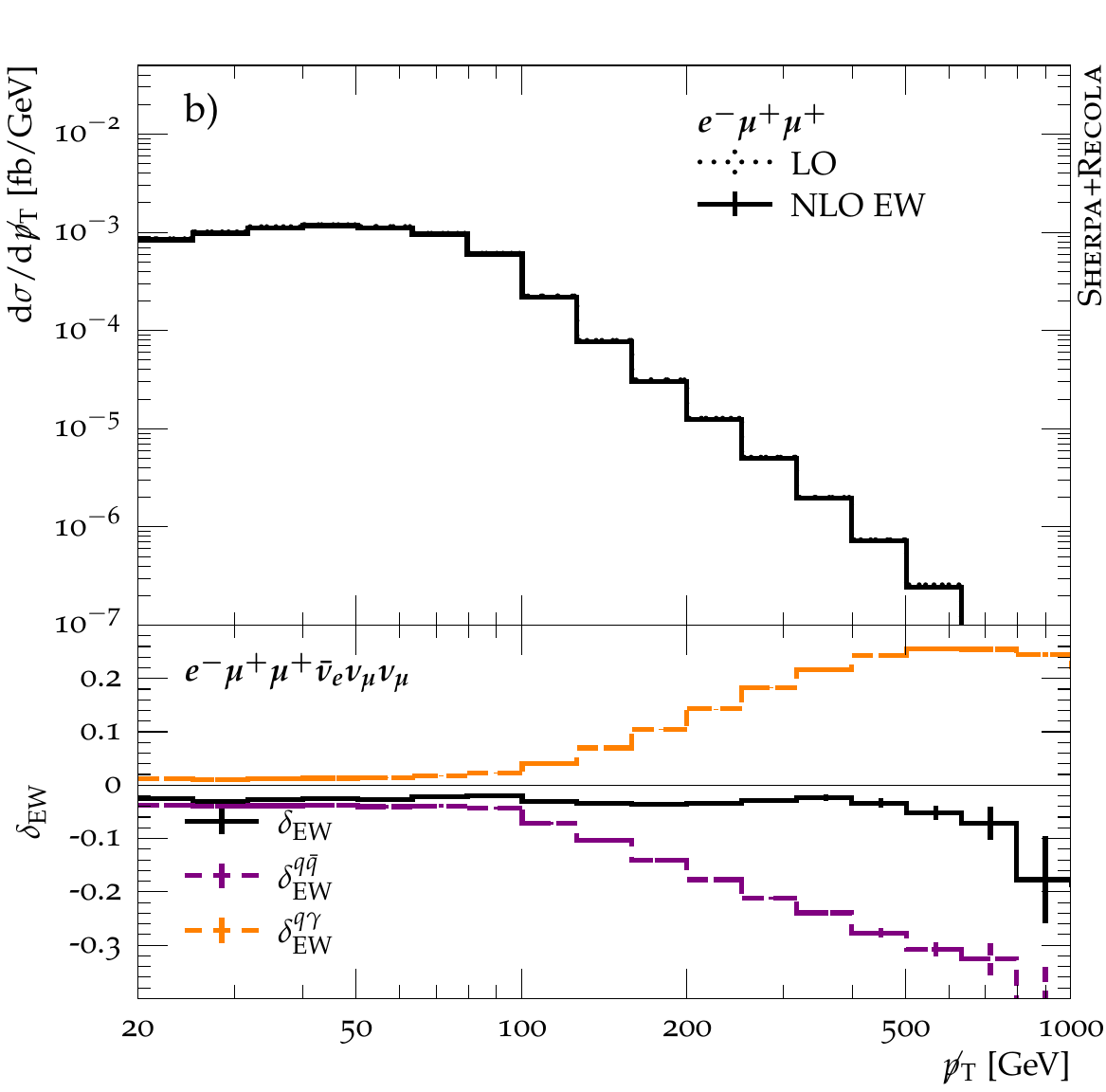}
    \includegraphics[width=\textwidth]{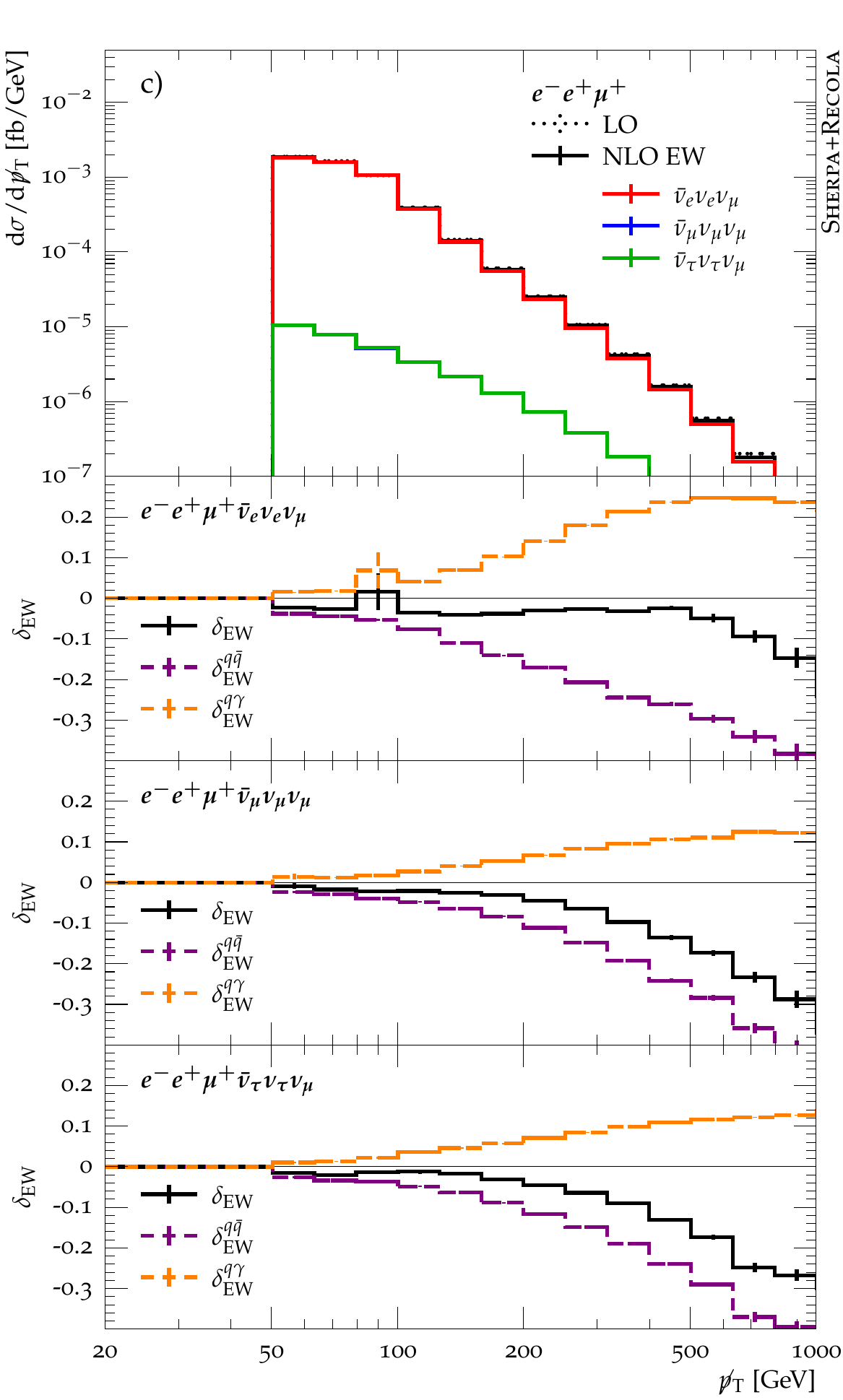}
  \end{minipage}
  \caption{
    Electroweak corrections to the missing transverse momentum distribution. 
    \label{fig:MET}
  }
\end{figure}

The electroweak corrections for the missing transverse 
momentum distribution is displayed in Fig.\ \ref{fig:MET}. 
Throughout the spectrum, the contribution of the 1 SFOS lepton pair 
channel remains approximately constant, while the 2 SFOS lepton pair
slowly increases in importance. 
The photon-induced corrections increase to maximum of $+25\%$ 
at $\pTmis\approx 500\,\text{GeV}$ and before decreasing rapidly 
thereafter. 
Up until that point they, to a very good degree, cancel the 
genuine electroweak corrections in the $q\bar{q}$ channel, 
resulting in an almost constant and unnaturally small total \deltaEW. 
Only thereafter the electroweak correction rises in the familiar 
fashion. 
The \deltaEWqq\ on its own, shows this behaviour in the 
whole range $\pTmis>100\,\text{GeV}$, as expected. 
In the subdominant lepton channels with only $WZZ$ topologies, 
where the photon-induced compensation is much smaller, 
the Sudakov-like shape of the electroweak corrections 
is much more apparent.

\begin{figure}[p]
  \centering
  \begin{minipage}{0.47\textwidth}
    \includegraphics[width=\textwidth]{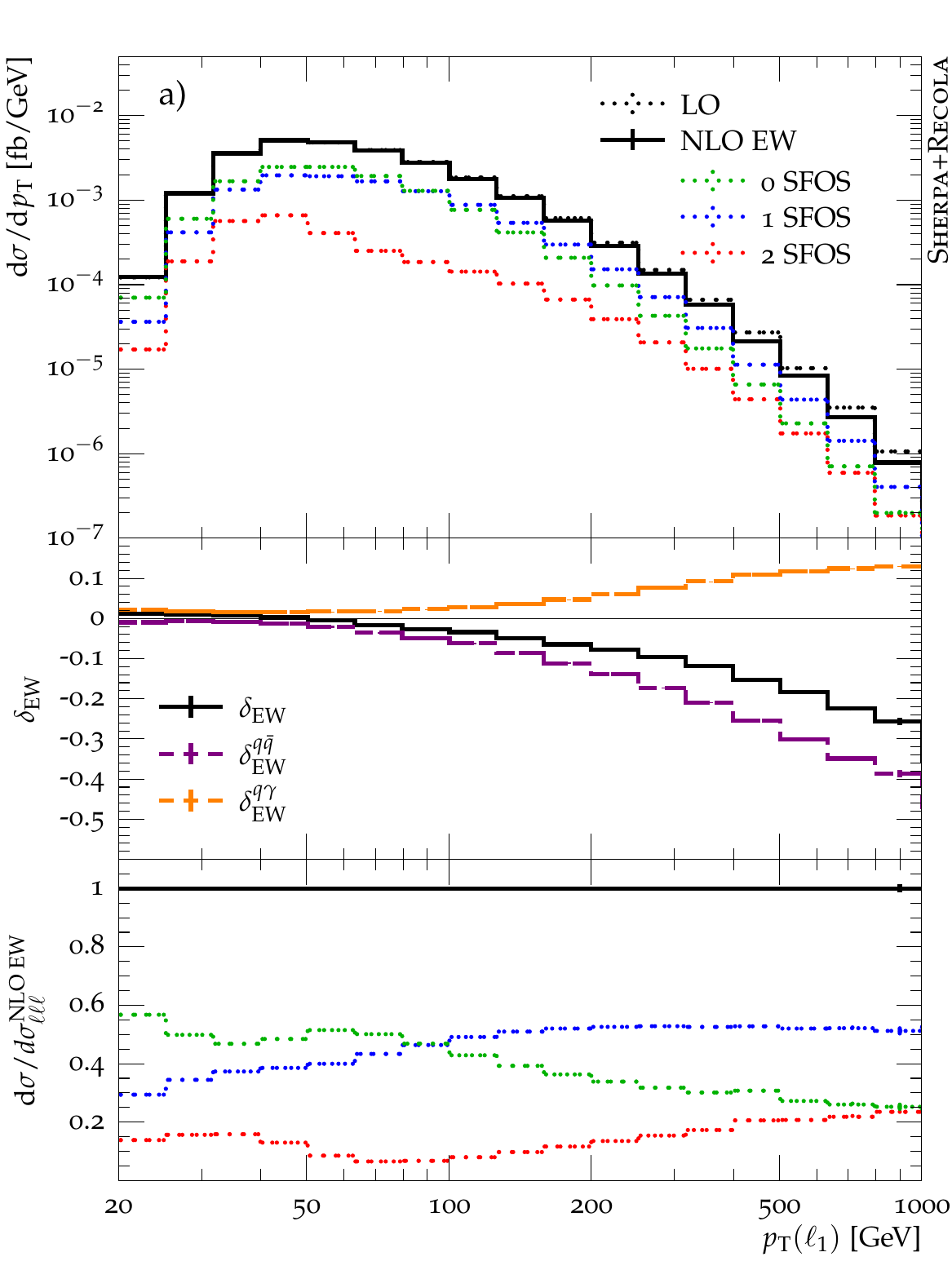}
    \includegraphics[width=\textwidth]{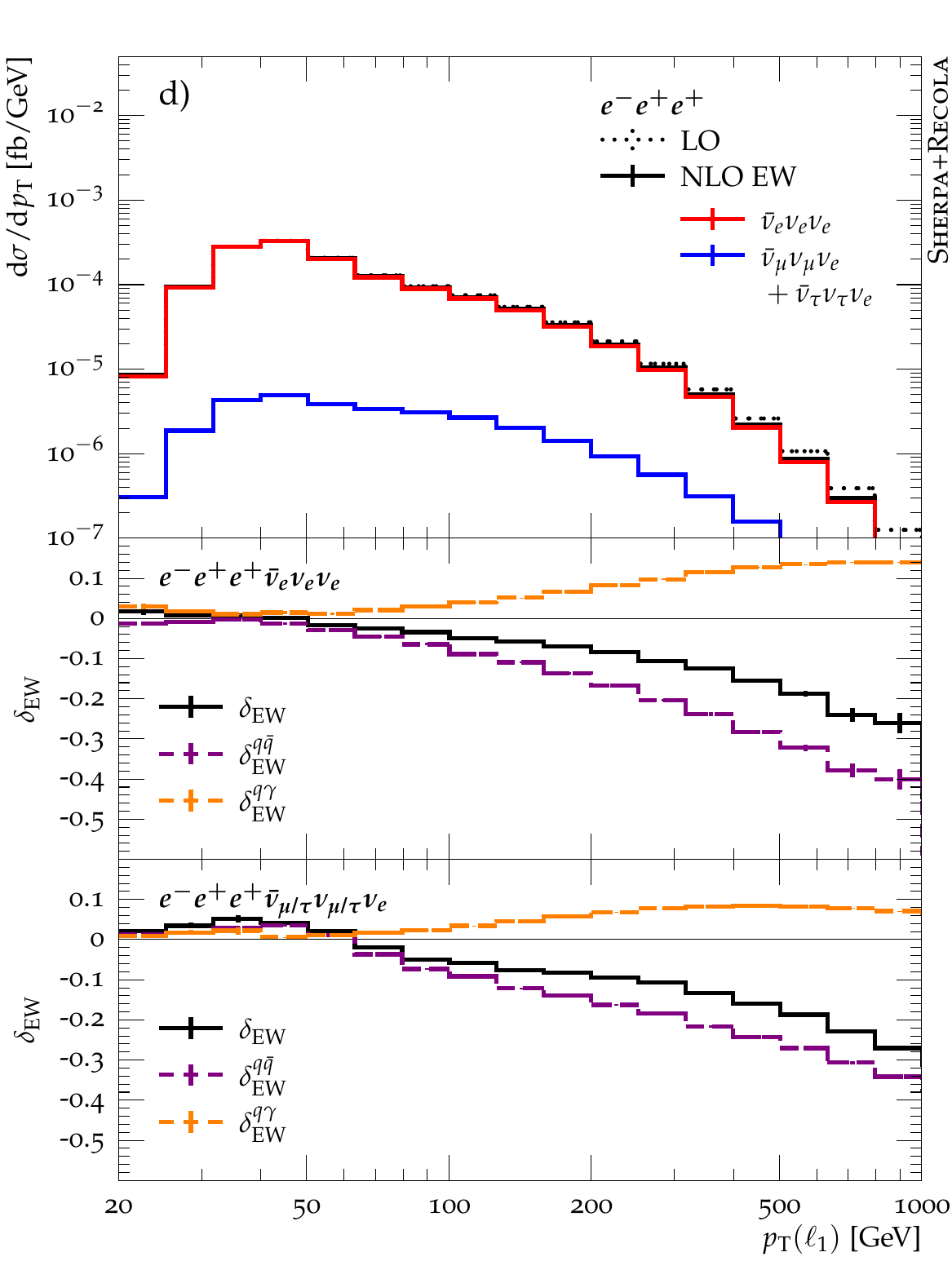}
  \end{minipage}
  \hfill
  \begin{minipage}{0.47\textwidth}
    \includegraphics[width=\textwidth]{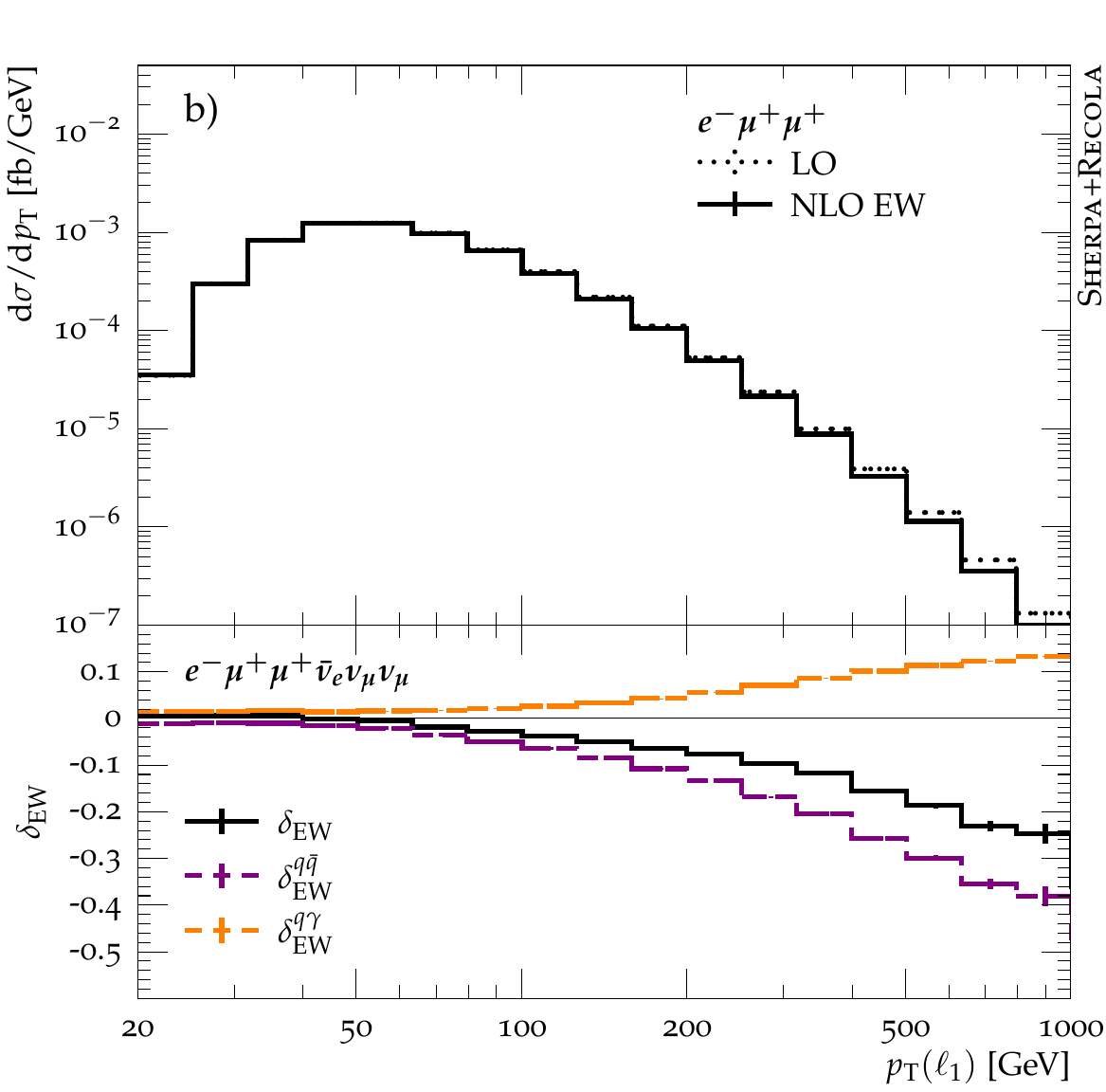}
    \includegraphics[width=\textwidth]{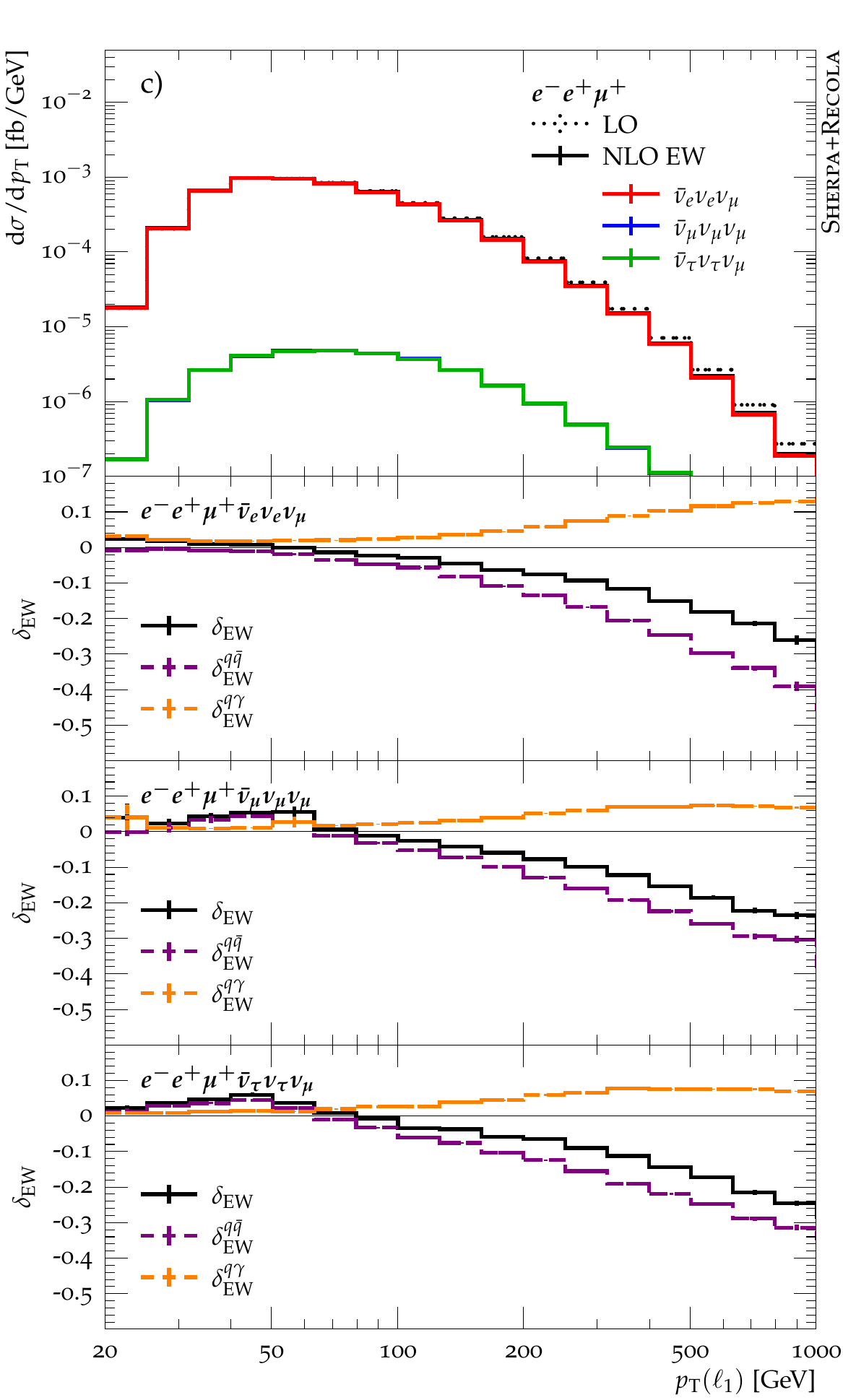}
  \end{minipage}
  \caption{
    Electroweak corrections to the leading lepton transverse momentum distribution. 
    \label{fig:pT_l1}
  }
\end{figure}

\begin{figure}[p]
  \centering
  \begin{minipage}{0.47\textwidth}
    \includegraphics[width=\textwidth]{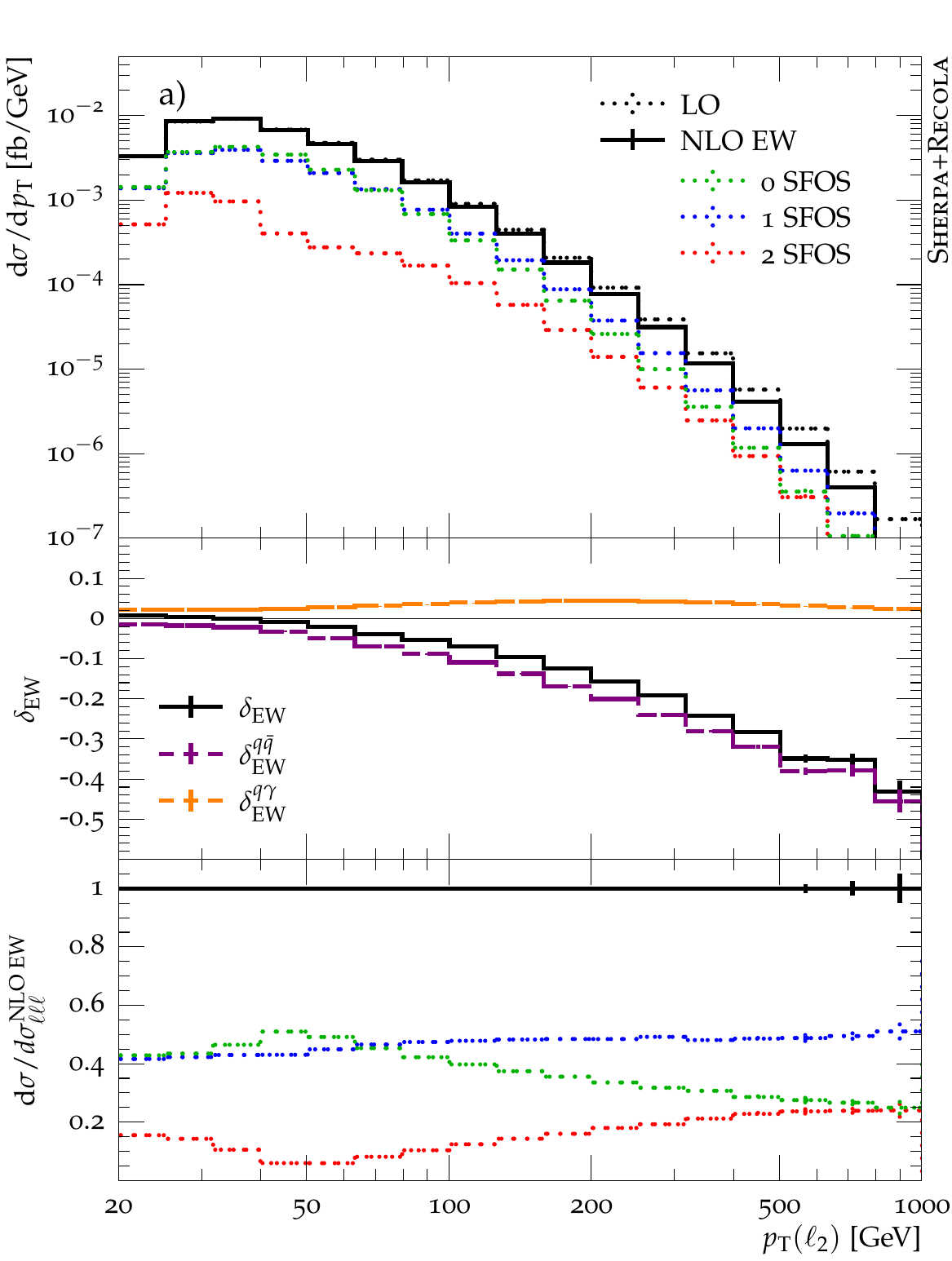}
    \includegraphics[width=\textwidth]{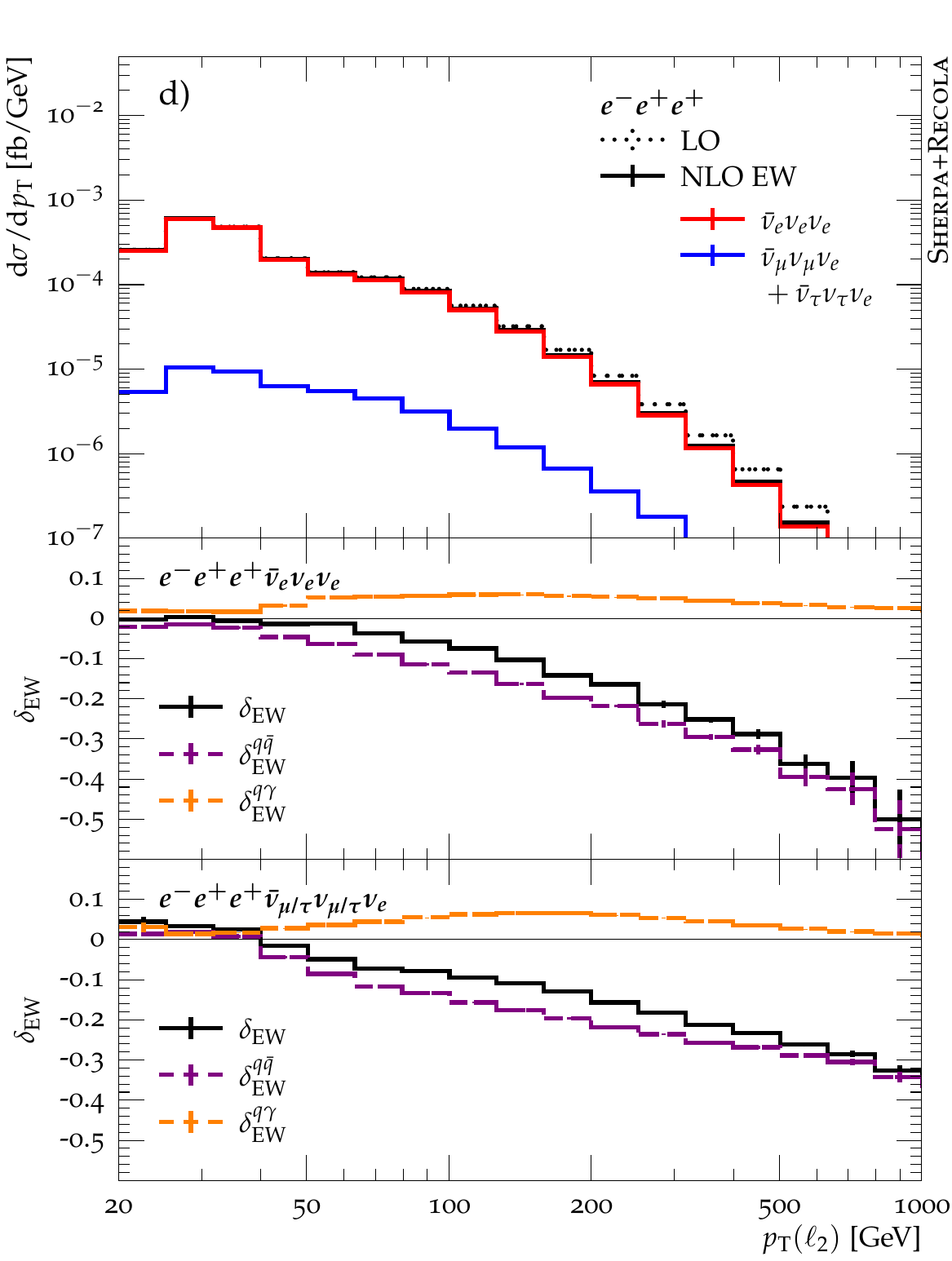}
  \end{minipage}
  \hfill
  \begin{minipage}{0.47\textwidth}
    \includegraphics[width=\textwidth]{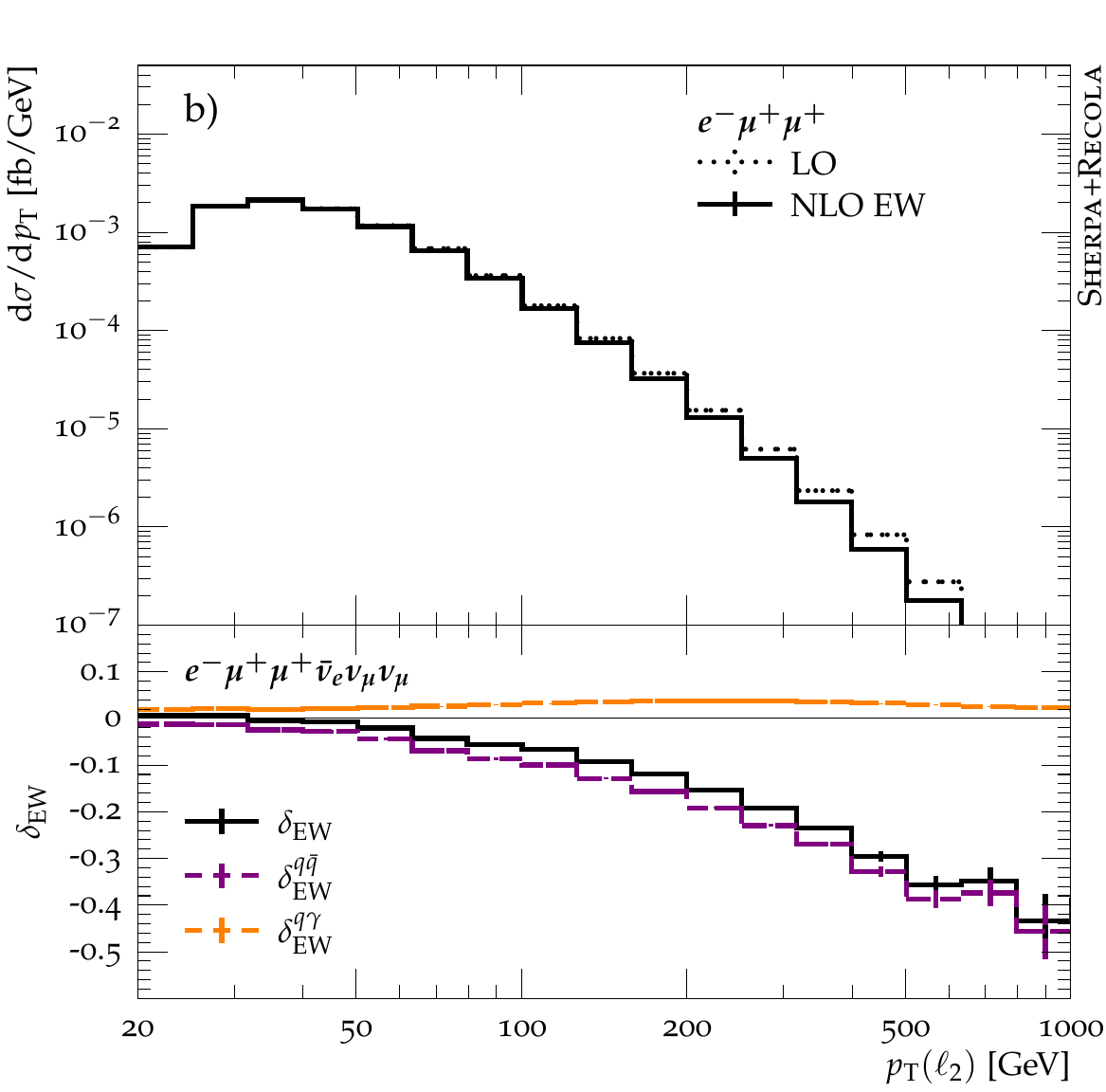}
    \includegraphics[width=\textwidth]{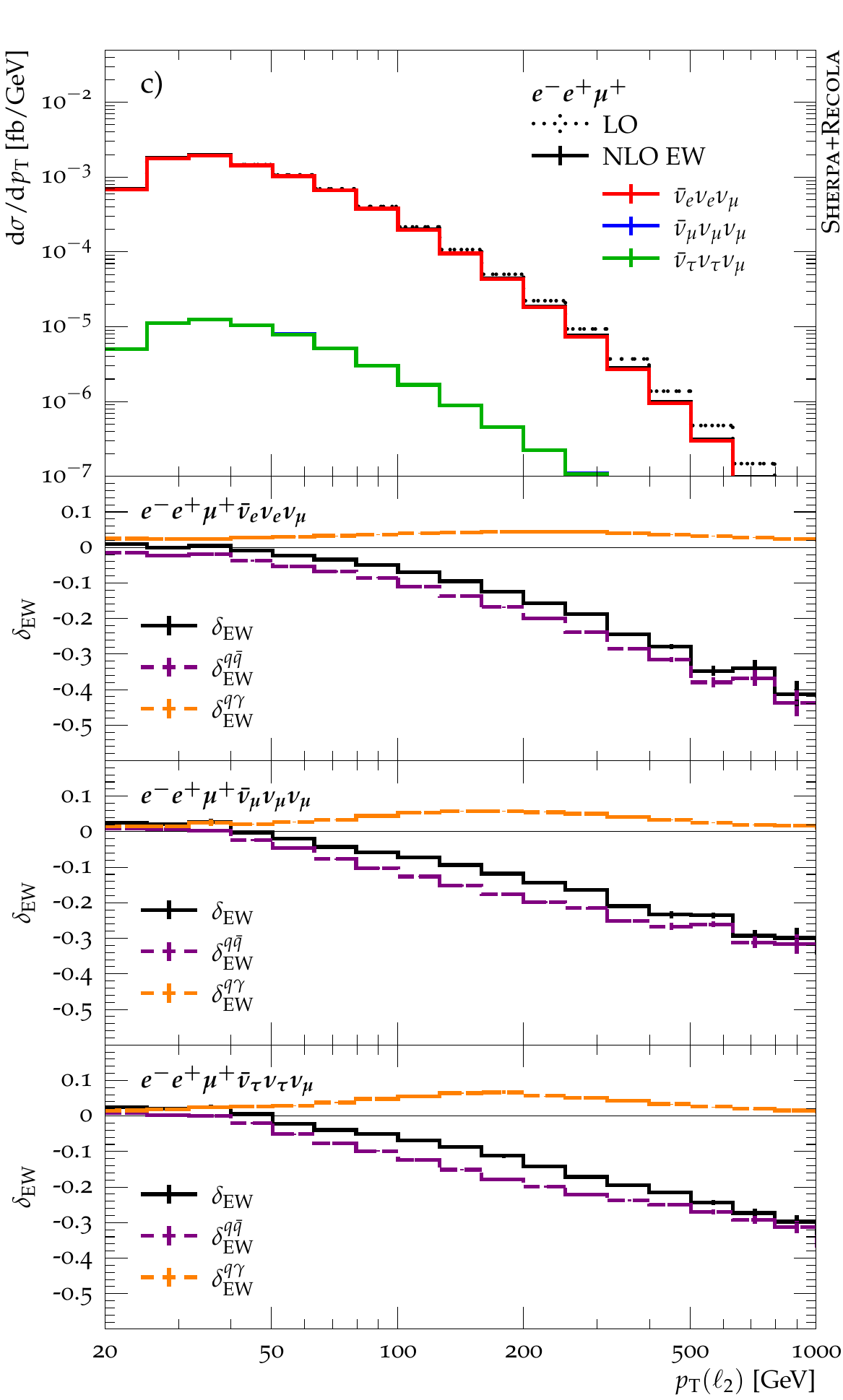}
  \end{minipage}
  \caption{
    Electroweak corrections to the subleading lepton transverse momentum distribution. 
    \label{fig:pT_l2}
  }
\end{figure}

\begin{figure}[p]
  \centering
  \begin{minipage}{0.47\textwidth}
    \includegraphics[width=\textwidth]{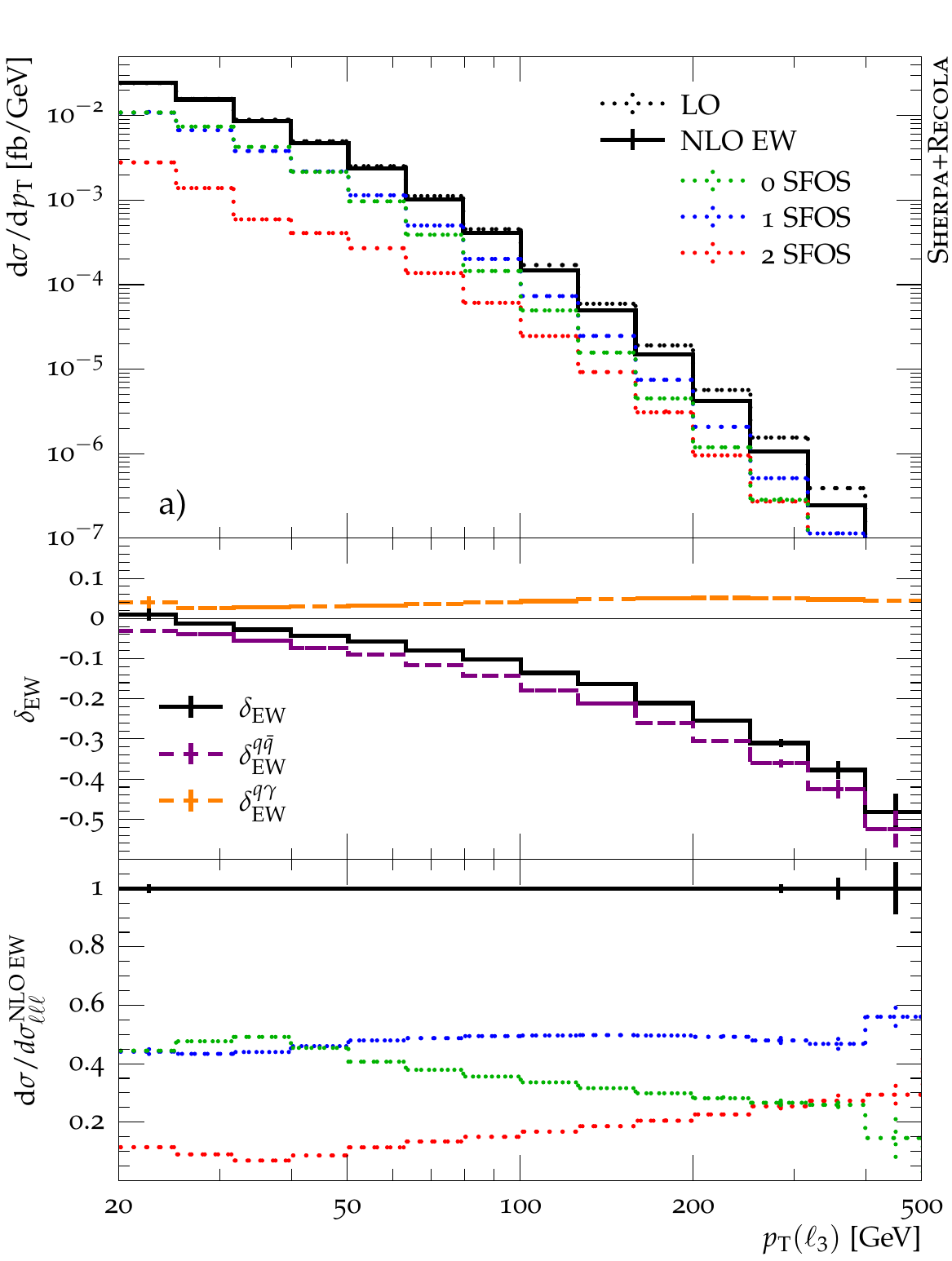}
    \includegraphics[width=\textwidth]{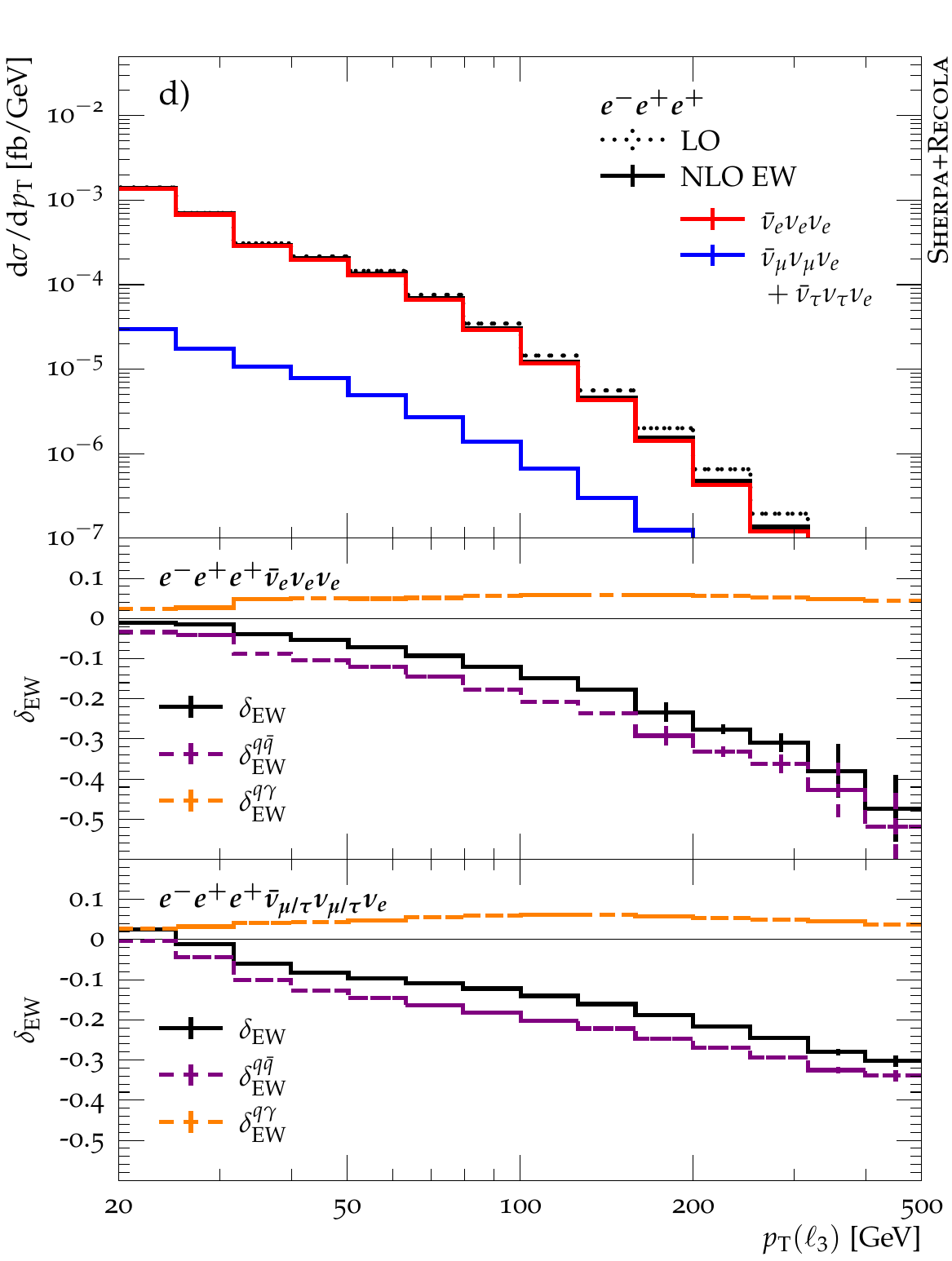}
  \end{minipage}
  \hfill
  \begin{minipage}{0.47\textwidth}
    \includegraphics[width=\textwidth]{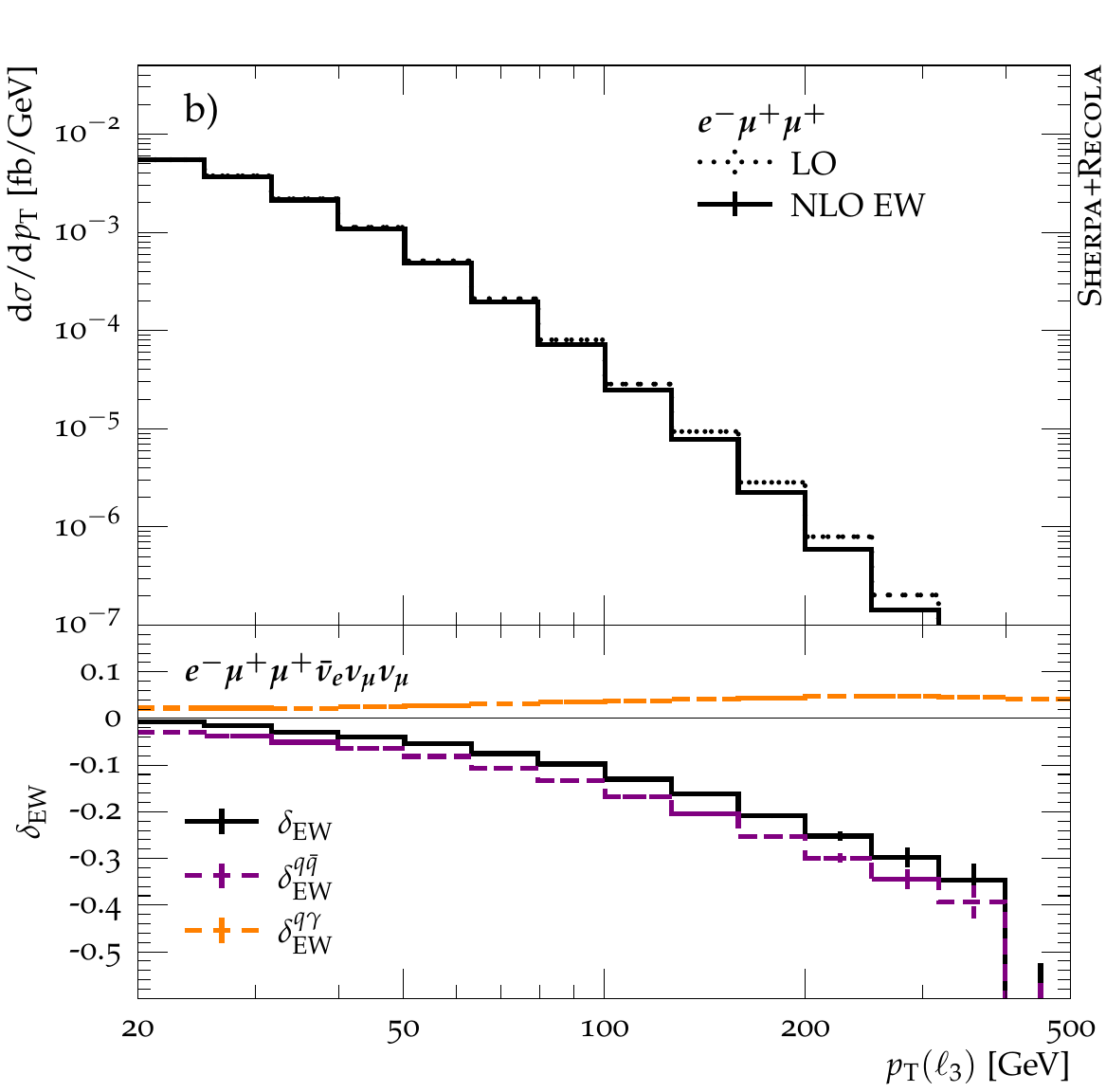}
    \includegraphics[width=\textwidth]{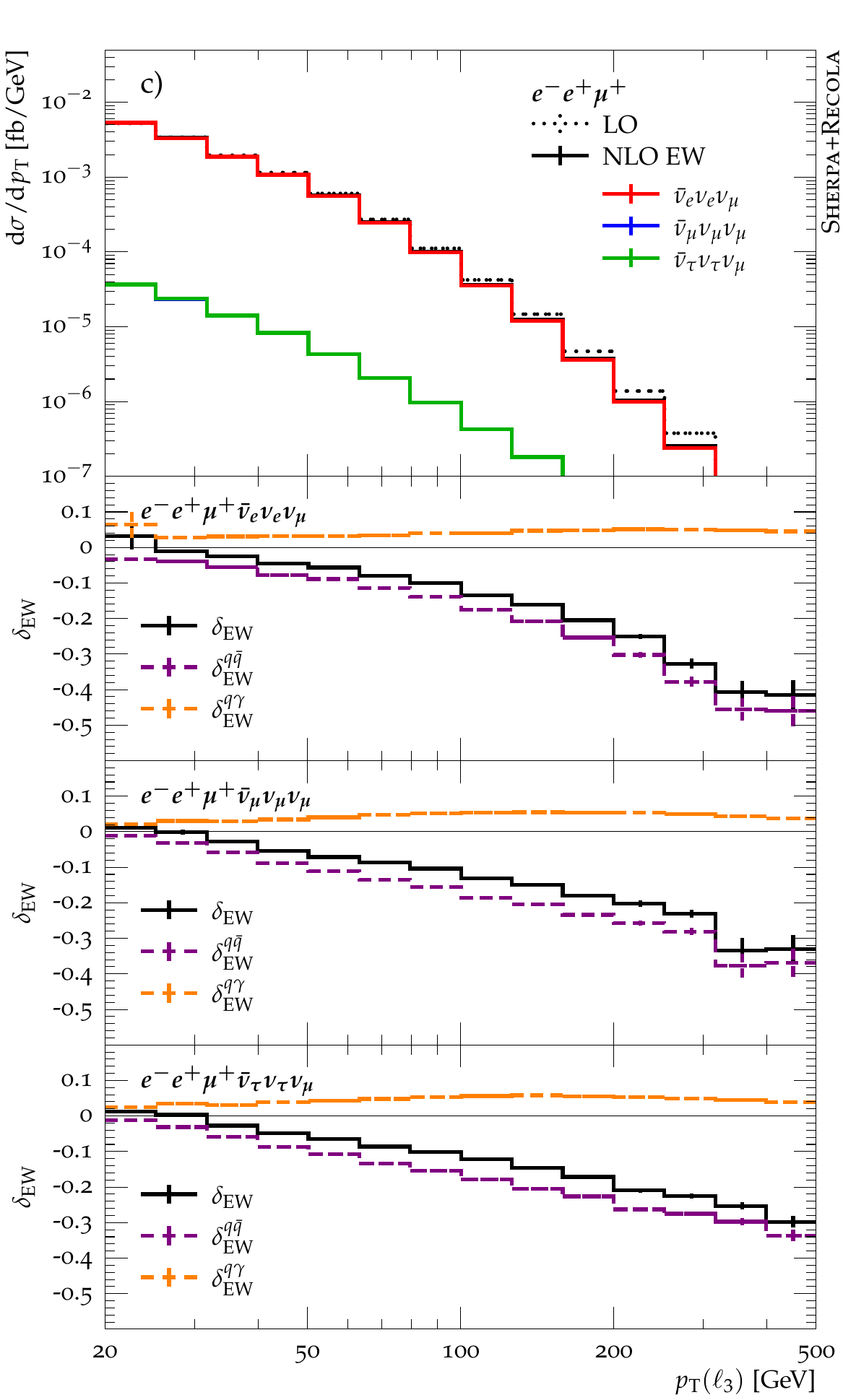}
  \end{minipage}
  \caption{
    Electroweak corrections to the third-leading lepton transverse momentum distribution. 
    \label{fig:pT_l3}
  }
\end{figure}

Fig.\ \ref{fig:pT_l1}--\ref{fig:pT_l3} finally display the 
transverse momentum spectra of all three leptons, sorted by \pT. 
The general picture is similar for all of them. 
The photon-induced real emission corrections are smaller than in the 
missing transverse momentum case and the genuine electroweak corrections 
in the $q\bar{q}$-channel is larger, resulting in a not-substantially 
disturbed shapes of the total electroweak corrections. 
At transverse momenta of 500\,GeV they amount to 
$\deltaEWqq\approx -30\%/-40\%/-50\%$ and 
$\deltaEW\approx -20\%/-35\%/-45\%$ for 
the first/second/third leading lepton.

In the case of the leading lepton they turn positive 
for very small transverse momenta, below 40\,GeV. 
In the pure $WZZ$ channels this positive contribution 
is somewhat larger and extends to slightly higher 
transverse momenta.

\section{Conclusions}
\label{sec:conclusions}

In this paper we have calculated the next-to-leading order 
electroweak corrections to off-shell $W^-W^+W^+$ production, 
namely to trilepton 
$\ell_1^-\ell_2^+\ell_3^+\bar{\nu}_{\ell_1}\nu_{\ell_2}\nu_{\ell_3}$ 
($\ell_i=e,\mu$) signatures. 
All triple, double, single and non resonant topologies and 
interferences of diagrams with all different vector boson 
($W,Z,\gamma$) intermediate states are included. 

We have confirmed that the electroweak corrections exhibit 
substantial \emph{accidental} cancellations between 
genuine (electro)weak corrections, dominated by the exchange 
of virtual electroweak gauge bosons, and the photon-induced 
real emission corrections that feature an additional jet in 
the final state, first observed in \cite{Dittmaier:2017bnh}. 
The resulting next-to-leading order electroweak corrections 
amount to approximately $-2.0\%$ ($-5.2\%$ genuine 
(electro)weak corrections in the $q\bar{q}$ channel and 
$+3.2\%$ in the photon-induced jet radiation channel) 
for the inclusive fiducial 
cross section with the definition of the fiducial region 
defined in Tab.\ \ref{tab:cuts}, which includes a moderate 
jet veto. 
It needs to be stressed that the precise impact of the 
strictly positive contribution from the photon-induced 
corrections strongly depends on the precise form and value 
of this jet veto. 
The electroweak corrections increase rapidly if either either 
the trilepton invariant mass, the missing transverse momentum 
or any of the lepton transverse momenta are increased. 
For trilepton invariant masses larger than 500\,GeV they increase 
to about $-7.7\%$ ($-16.3\%+8.6\%$). 
Similarly, for missing transverse momenta larger than 200\,GeV 
the complete electroweak corrections amount to $-3.4\%$ 
($-20.7\%+17.3\%$). 
The aforementioned compensation of genuine (electro)weak 
corrections and photon-induced jet radiation contributions 
was found to strongly depend on the observable studied. 
This further emphasises the necessity to compute either 
contribution exactly. 

Further, due to the fully off-shell nature of this calculation, 
the electroweak corrections in regions like $m_{3\ell}<3\,m_W$ 
has now been calculated for the first time. 
In addition to $\gamma\gamma\gamma$, $\gamma\gamma W$ and 
$\gamma\gamma Z$ \cite{Greiner:2017mft} $WWW$ is now the fourth 
triboson process known to NLO QCD and NLO EW accuracy in the 
fully off-shell case.
Finally, it is worth to stress, that neither of these individually 
large effects, genuine (electro)weak corrections in the $q\bar{q}$ 
channel as well as photon-induced jet radiation, is incorporated 
in any Monte-Carlo event generator in use by the experiments. 
The presented results detail, that they must be included together 
consistently as $\order(\alpha)$ corrections to the inclusive 
process, otherwise important and far-reaching cancellations are 
missed.

\subsection*{Acknowledgements}

M.S.\ would like to thank M.\ Pellen and S.\ Br\"auer 
for help and clarifications on the use of \Recola and 
its interface to \Sherpa.
This work has received funding from the European Union's 
Horizon 2020 research and innovation programme as part of 
the Marie Sklodowska-Curie Innovative Training Network 
MCnetITN3 (grant agreement no.\ 722104).

\bibliographystyle{amsunsrt_modpp}
\bibliography{journal}
  \end{document}